\documentclass[aps,prd,groupedaddress,eqsecnum]{revtex4}

\usepackage{graphicx}
\usepackage{amssymb}
\usepackage{color}
\usepackage{subfigure}

\def\C{\mathbb C}
\def\Z{\mathbb Z}
\def\R{\mathbb R}
\def\N{\mathbb N}

\begin{document}

\title{The Bogoliubov-de Gennes system, the AKNS hierarchy,
and nonlinear quantum mechanical supersymmetry}

\author{Francisco Correa$^{\dagger}$, Gerald V.~Dunne$^{\ddagger}$ and Mikhail S. Plyushchay$^{\dagger}$}

\affiliation{$^{\dagger}$Departamento de F\'{\i}sica, Universidad de
Santiago de Chile, Casilla 307, Santiago 2, Chile\\ $^{\ddagger}$Physics Department,
University of Connecticut, Storrs CT 06269, USA}

\begin{abstract}
We show that the Ginzburg-Landau expansion of the grand potential
for the Bogoliubov-de Gennes Hamiltonian is determined by the
integrable nonlinear equations of the AKNS hierarchy, and that this
provides the natural mathematical framework for a hidden nonlinear
quantum mechanical supersymmetry underlying the dynamics.

\end{abstract}

\maketitle

\section{ Introduction: The Bogoliubov-de Gennes Hamiltonian }


The Bogoliubov-de Gennes (BdG) Hamiltonian \cite{degennes}  is
\begin{equation}
    H=
    \begin{pmatrix}
    {-i\frac{d}{dx}&\Delta(x)\cr \Delta^*(x) & i\frac{d}{dx}}
    \end{pmatrix}\quad ,
    \label{ham-bdg}
\end{equation}
where the (in general complex) function $\Delta(x)$ plays the role
of an order parameter, a condensate, or a gap function, depending on
the physical context. The BdG Hamiltonian is relevant for the
physics of: 
(i)  crystalline condensates \cite{thies-gn,bd2,bdt} in the chiral Gross-Neveu and Nambu-Jona Lasinio models \cite{gross,dhn,pap,klimenko,feinberg}; 
(ii) fractional fermion number \cite{jackiw,niemi}; 
(iii) the Peierls effect \cite{peierls,belokolos}; 
(iv) polaron crystals in conducting polymers
\cite{braz,horovitz,campbell,heeger}; 
(v) the pair potential in inhomogeneous quasi-1D superconductors \cite{eilenberger,kuper,buzdin,leggett}; 
(vi) the order parameter for superconductors in a ferromagnetic field \cite{machida}; 
(vii) incommensurate charge-density waves \cite{sakita,mertsching}; 
(viii) phase structure of ultracold atomic gases \cite{pitaevskii}.
In this paper we discuss the rich mathematical physics properties of the BdG Hamiltonian and in particular its close relation to nonlinear quantum mechanical supersymmetry \cite{andrianov,klishevich}.

An important object for the analysis of physical models based on the
BdG Hamiltonian is the Gorkov resolvent
\cite{gorkov,dickey,das,gesztesy}:
\begin{equation}
    R(x; E)\equiv \langle x| \frac{1}{H-E} |x\rangle \quad .
    \label{res}
\end{equation}
The resolvent (\ref{res}) is clearly a $2\times 2$ matrix. It
encodes the spectral properties of the BdG Hamiltonian. For example,
the spectral function (density of states) characterizing the
single-particle spectrum of fermions in the presence of the
condensate $\Delta(x)$ is
\begin{equation}
    \rho(E)=\frac{1}{\pi}{\rm Im}\,{\rm Tr}_{D,x}\left[R(x;E+i\epsilon)\right]\quad ,
    \label{spectral}
\end{equation}
where the trace is a Dirac trace as well as a spatial trace. The local density of
states is defined without the spatial trace:
\begin{equation}
    \rho(x; E)=\frac{1}{\pi}{\rm Im}\,{\rm Tr}_{D}\left[R(x;E+i\epsilon)\right]\quad .
    \label{local-dos}
\end{equation}
The thermodynamic grand potential density is defined as
\begin{equation}
    \psi(x)=-\frac{1}{\beta} \int_{-\infty}^\infty  dE\,\rho(x; E)\,
    \ln\left(1+e^{-\beta(E-\mu)}\right)\quad, \label{grand}
\end{equation}
where $\beta=\frac{1}{T}$ is the inverse temperature, and $\mu$ is
the chemical potential. The grand potential is given by the
spatial integral
\begin{eqnarray}
    \Psi\equiv \int dx\, \psi(x) \quad ,
    \label{grand-pot}
\end{eqnarray}
and the thermodynamic properties [entropy, density, free energy,
\dots]  of the system can be derived from $\Psi$ \cite{kapusta}. All information
concerning the condensate $\Delta(x)$ is contained in the local
density of states $\rho(x; E)$, and hence in the resolvent $R(x;
E)$.

For a one-dimensional system [as is being studied here], the
resolvent can be expressed in Wronski form, using two independent
solutions of $H\psi=E \psi$. By elementary arguments
\cite{dickey,waxman,leggett,bd2}, this construction implies that $R(x; E)$
must satisfy a $2\times 2$ matrix first-order differential
equation, known as the Dikii-Eilenberger equation:
\begin{eqnarray}
    \frac{\partial}{\partial x}R(x; E)\, \sigma_3
    &=&i\, \left[
    \begin{pmatrix}
    {E&-\Delta(x) \cr
    \Delta^*(x) & -E}
    \end{pmatrix}, R(x; E)\,\sigma_3
    \right]\quad .
    \label{dikii}
\end{eqnarray}
It must also satisfy reality and normalization conditions [expressed here for $E$
not in the spectrum, with suitable extensions to other energy branches]:
\begin{eqnarray}
    \det R=-\frac{1}{4} \quad , \qquad R^\dagger =R\quad .
    \label{conditions}
\end{eqnarray}
For example, if $\Delta=M$ is a constant, then we have the familiar solution
\begin{eqnarray}
    R(x; E)=
    \frac{1}{2\sqrt{M^2-E^2}}
    \begin{pmatrix}
    {E& M\cr
    M&E}
    \end{pmatrix}
    \label{constant-res}
\end{eqnarray}
and it is straightforward to check that this resolvent satisfies the
Dikii-Eilenberger equation (\ref{dikii}).

For non-trivial $\Delta(x)$ it is not so easy to find the exact resolvent,
so various expansions have been developed. One such expansion is based
on a large $E$ asymtotic expansion of the resolvent:
\begin{eqnarray}
    R(x; E)=\frac{1}{2}\sum_{n=0}^\infty\frac{\hat{r}_n(x)}{E^n}\quad.
    \label{r-expansion}
\end{eqnarray}
The energy integrals in (\ref{grand}) and (\ref{grand-pot}) can
now be done, leading to an expansion of the
grand potential density and the grand potential
\begin{eqnarray}
    \psi(x)&=&\sum_{n=0}^\infty \alpha_n(T, \mu)\, \hat{g}_n(x)\quad,\\
    \Psi&=&\sum_{n=0}^\infty \alpha_n(T, \mu)\, \hat{G}_n \quad ,
    \qquad \hat{G}_n\equiv \int dx\, \hat{g}_n(x)\quad .
\end{eqnarray}
This expansion generalizes a result of Perelomov  \cite{perelomov} for the grand potential of systems with Hamiltonians of Schr\"odinger form, rather than of Dirac form, and
which is related in turn to the KdV hierarchy. On the other hand, we show in the next Section that
the Dirac case is naturally related to the AKNS hierarchy.
The $T$ and $\mu$ dependent coefficients $\alpha_n(T, \mu)$ are
known functions \cite{thies-gn,bd2}. The $\hat{g}_n(x)$  come from
the diagonal elements of  the $2\times 2$ matrices $\hat{r}_n(x)$
in the resolvent expansion (\ref{r-expansion}):
\begin{eqnarray}
    \hat{r}_n(x)=
    \begin{pmatrix}
    {\hat{g}_n(x) & \hat{f}_{n-1}(x)\cr
    \hat{f}_{n-1}^*(x) & \hat{g}_n(x)}
    \end{pmatrix}\quad.
    \label{rexpmatrix}
\end{eqnarray}
Substituting the expansion (\ref{r-expansion}) into the
Dikii-Eilenberger equation (\ref{dikii}), we obtain the simple
recursion equations
\begin{eqnarray}
    \hat{f}_n&=&-\frac{i}{2}\hat{f}_{n-1}^{\,\prime}+\Delta \,
    \hat{g}_n\quad,\\
    \hat{g}_n^\prime &=& i\left(\Delta^*\hat{f}_{n-1}-\Delta\, \hat{f}_{n-1}^*
    \right)\quad,\\
    \hat{g}_0&=&1\quad,\\
    \hat{f}_{-1}&=&0\quad.
    \label{recursion}
\end{eqnarray}
The initial values follow from the $\Delta=M$ (constant)  case above. It is simple
to generate these terms recursively. The first few off-diagonal terms are:

\begin{eqnarray}
    \hat{f}_{-1}&=&0\quad, \nonumber\\
    \hat{f}_0&=&\Delta\quad, \nonumber\\
    \hat{f}_1&=& -\frac{i}{2}\Delta^\prime\quad,\nonumber\\
    \hat{f}_2&=& -\frac{1}{4}\left(\Delta^{\prime\prime}
    -2|\Delta|^2\Delta \right)\quad,\nonumber\\
    \hat{f}_3&=& \frac{i}{8}\left(\Delta^{\prime\prime\prime}
    -6|\Delta|^2\Delta^\prime\right)\quad, \nonumber\\
    \hat{f}_4&=&\frac{1}{16}\left(\Delta^{(\text{iv})}-8|\Delta|^2
    \Delta^{\prime \prime}-2 \Delta^2
    \Delta^{ * \prime \prime}-4|\Delta^{\prime}|^2 \Delta-6
    \Delta^*  \Delta^{\prime 2}+6|\Delta|^4 \Delta
    \right)\quad, \nonumber \\
    \hat{f}_5&=& -\frac{i}{32}\left(\Delta^{(\text{v})} -10|\Delta|^2\Delta^{\prime \prime
    \prime} -10(\Delta \Delta^{* \prime}+2\Delta^*\Delta^{\prime})\Delta^{\prime
    \prime}-10\Delta \Delta^{\prime}\Delta^{* \prime \prime}-10\Delta^{* \prime}
    \Delta^{\prime 2}+30|\Delta|^4\Delta^{\prime}
    \right)\quad,\nonumber\\
    &\vdots&
    \end{eqnarray}
The first few diagonal terms are:
\begin{eqnarray}
    \hat{g}_0&=&1\quad, \nonumber\\
    \hat{g}_1&=&0\quad, \nonumber\\
    \hat{g}_2&=&\frac{1}{2}|\Delta|^2\quad, \nonumber\\
    \hat{g}_3&=& \frac{i}{4}\left(\Delta\Delta^{\prime\,*}
    -\Delta^\prime\Delta^*\right)\quad,\nonumber\\
    \hat{g}_4&=& \frac{1}{8}\left(3|\Delta|^4+3|\Delta^\prime|^2
    -\left(|\Delta|^2\right)^{\prime\prime}\right)\quad,
    \nonumber\\
    \hat{g}_5&=&\frac{i}{16}\left(   \Delta''' \Delta^* - \Delta
    \Delta^{* \prime \prime \prime} +\Delta' \Delta^{* \prime \prime}-
    \Delta^{ \prime \prime} \Delta^{* \prime} +6|\Delta|^2(\Delta^{* \prime} \Delta-\Delta'\Delta^*)
     \right)\quad,\nonumber \\
     \hat{g}_6&=&\frac{1}{32}\left(  \Delta^{(\text{iv})} \Delta^*+
     \Delta^{* (\text{iv})}\Delta -(|\Delta'|^2)''+3|\Delta''|^2 -10|
     \Delta|^2(\Delta''\Delta^*+\Delta^{* \prime \prime}\Delta)-
     5(\Delta^{*2}\Delta^{\prime 2}+\Delta^{2}\Delta^{ * \prime 2})+10|\Delta|^6
     \right)\nonumber \\
     &=& \frac{1}{32}\left( (|\Delta|^2)^{(\text{iv})}-5(|\Delta'|^2)'' +
     5 |\Delta ''|^2-10 |\Delta|^2(\Delta''\Delta^*+\Delta^{* \prime \prime}\Delta)-
     5(\Delta^{*2}\Delta^{\prime 2}+\Delta^{2}\Delta^{ * \prime 2})+10|\Delta|^6
     \right)\quad, \nonumber \\
    &\vdots&
\end{eqnarray}

Note that in principle there are integration constants appearing
in the $\hat{g}_n$, but these all vanish by virtue of the
normalization condition (\ref{conditions}).

These recursively generated  quantities $\hat{f}_n$ and $\hat{g}_n$  have many remarkable
properties, as is discussed in this paper. For example, the
functional variation of $\hat{G}_n$ is related to
$\hat{f}_{n-2}$ as:
\begin{eqnarray}
     \frac{\delta \hat{G}_n}{\delta \Delta^*(x)}=\frac{(n-1)}{2}
     \,\hat{f}_{n-2}(x)\quad.
     \label{relation}
\end{eqnarray}
The physical reason for this remarkable relation is the following.
The variation of the grand potential (\ref{grand-pot}) can clearly
be expressed as
\begin{eqnarray}
    \frac{\delta \Psi}{\delta \Delta^*(x)}=\sum_{n=0}^\infty \alpha_n(T, \mu)\,
    \frac{\delta \hat{G}_n}{\delta \Delta^*(x)}\quad.
    \label{var-1}
\end{eqnarray}
On the other hand, the grand potential (\ref{grand-pot}) can
alternatively be written \cite{bd2}, including the
thermodynamical Fermi factor in the energy trace, as
\begin{eqnarray}
    \Psi=\int\frac{dE}{2\pi} \frac{1}{e^{\beta(E-\mu)}+1}\,{\rm Tr}_{D, x}\, \ln\,
    \left[\gamma^0\left(E-H\right)\right]\quad.
\end{eqnarray}
We choose to work in the Dirac gamma matrix representation
\begin{eqnarray}
    \gamma^0=\sigma_1\equiv \begin{pmatrix}{0&1\cr 1&0}\end{pmatrix}
    \quad , \qquad \gamma^1=-i\sigma_2\equiv \begin{pmatrix}{0&-1\cr
    1&0}\end{pmatrix} \quad , \qquad \gamma^5=\sigma_3\equiv
    \begin{pmatrix}{1&0 \cr 0&-1}\end{pmatrix}\quad. \label{gamma-1}
\end{eqnarray}
Then the BdG Hamiltonian $H$ can be expressed as
\begin{equation}
    H=-i\gamma^5 \frac{d}{dx}+\gamma^0 \left[\frac{1}{2}
    \left(1-\gamma^5\right)\Delta(x)+
    \frac{1}{2}\left(1+\gamma^5\right)\Delta^*(x)\right]\quad.
\label{ham}
\end{equation}
Therefore, the variation of the grand potential with respect to
$\Delta^*$ can be written as
\begin{eqnarray}
    \frac{\delta \Psi}{\delta \Delta^*(x)}&=&\int\frac{dE}{2\pi} \frac{1}{e^{\beta(E-\mu)}+1}
    \,{\rm tr}_{D}\left[\gamma^0({\bf 1}+\gamma^5)\, R(x; E)\right] \nonumber\\
    &=&\sum_{n=2}^\infty \alpha_n(T, \mu)\,
    \frac{(n-1)}{2}\hat{f}_{n-2}(x)\quad.
    \label{var-2}
\end{eqnarray}
Comparing (\ref{var-1}) and (\ref{var-2}) we find the non-trivial
relation (\ref{relation}).

\section{AKNS hierarchy, finite-gap systems and hidden nonlinear supersymmetry}

The resolvent expansion functions $\hat{g}_n(x)$ and
$\hat{f}_n(x)$ found in the previous section form the backbone of
the (stationary, defocussing) AKNS hierarchy  \cite{akns, gesztesy, dickey, das}, a
hierarchy of integrable nonlinear differential equations. The idea
behind the integrability of the AKNS hierarchy can be expressed in
terms of a Lax pair \cite{lax},  $S$  and $H$, such that
\begin{eqnarray}
    [S, H]=0\quad.
\end{eqnarray}
There exists an infinite number of such $S$ commuting with $H$.
This is a statement of the integrability of the AKNS system, and
also of the BdG Hamiltonian. The Gorkov resolvent $R(x; E)$ is the
generating function for this infinite sequence of conserved
quantities. To see this explicitly, define the matrix differential
operator of order $(N+1)$ by the following polynomial expansion:
\begin{equation}
    \hat{S}_{N+1}=i \sum_{k=0}^{N+1}\hat{r}_{N+1-k}(x)\sigma_3
    H^k \quad ,\quad  N \in \mathbb{N}_0\quad .
\label{qnS}
\end{equation}
Here the $\hat{r}_k$ are the same quantities that appear in the
expansion (\ref{r-expansion}) of the Gorkov resolvent. Then the
recursion relations (\ref{recursion}) imply that
\begin{eqnarray}
    [\hat{r}_{N+1-k}\sigma_3, H]=
    2\begin{pmatrix}
    {0 & \hat{f}_{N+1-k}\cr
    -\hat{f}_{N+1-k}^*&0}
\end{pmatrix}
-2\begin{pmatrix}
    {0 & \hat{f}_{N-k}\cr
   -\hat{f}_{N-k}^*&0}
\end{pmatrix}\, H\quad .
\end{eqnarray}
This then implies that
\begin{equation}
    [\hat{S}_{N+1},H]=\begin{pmatrix}
    {0 &  2i\hat{f}_{N+1}(x) \cr  -2i \hat{f}^*_{N+1}(x) & 0}
\end{pmatrix}\quad .
\label{comm}
\end{equation}
The AKNS hierarchy is defined by considering finite linear
combinations of the    $\hat{S}_k$, with
coefficients $c_k$:
\begin{eqnarray}
    S_{N+1}\equiv \sum_{k=0}^{N+1} c_{N+1-k}\,
    \hat{S}_k\quad .
\label{q}
\end{eqnarray}
Then the $N^{\rm th}$ equation of the AKNS hierarchy is  $[S_{N+1},
H]=0$, which can equivalently be written as
\begin{eqnarray}
\sum_{k=0}^{N+1} c_{N+1-k}\, \hat{f}_k=0\quad .
\end{eqnarray}
For example, the first few equations in the hierarchy are
\begin{eqnarray}
    {\rm AKNS}_0 &:&\quad -\frac{i}{2}\Delta^\prime+c_1 \Delta=0\quad, \\
    {\rm AKNS}_1 &:&\quad   -\frac{1}{4}(\Delta^{\prime \prime}-2|\Delta|^2\Delta)
    -c_1\frac{i}{2}\Delta'+c_2 \Delta=0\quad,\\
    {\rm AKNS}_2 &:&\quad  \frac{i}{8}(\Delta^{\prime \prime \prime}-6|\Delta|^2\Delta^{\prime})
    -c_1\frac{1}{4}(\Delta^{\prime \prime}-2|\Delta|^2\Delta)-c_2\frac{i}{2}\Delta'+c_3 \Delta=0
    \quad, \\
    {\rm AKNS}_3 &:&\quad \frac{1}{16}\left(\Delta^{(\text{iv})}
    -8|\Delta|^2 \Delta^{\prime \prime}-2 \Delta^2  \Delta^{ * \prime \prime}
    -4|\Delta^{\prime}|^2 \Delta-6 \Delta^*  \Delta^{\prime 2}+6|\Delta|^4 \Delta
    \right) \nonumber \\ &&\hskip 2cm +
    c_1\frac{i}{8}(\Delta^{\prime \prime \prime}-6|\Delta|^2\Delta^{\prime})
    -c_2\frac{1}{4}(\Delta^{\prime \prime}-2|\Delta|^2\Delta)-c_3\frac{i}{2}\Delta'
    +c_4 \Delta=0\quad,\\
    \vdots &&\nonumber
\end{eqnarray}
The $N=1$ case, ${\rm AKNS}_1$, is the original AKNS equation,
which can also be called the complex nonlinear Schr\"odinger
equation.  Another way to see the relation between the Gorkov
resolvent and the AKNS hierarchy is to realize that the
Dikii-Eilenberger equation is a statement of the zero-curvature
representation of the AKNS hierarchy \cite{dickey,das}.

From a physical standpoint, the AKNS equations arise naturally in
a Ginzburg-Landau approach to the BdG system, in which one makes
an approximate expansion of the grand potential (\ref{grand-pot})
to some finite order, $N+1$:
\begin{eqnarray}
    \Psi_{\rm Ginzburg-Landau}=\sum_{k=0}^{N+1}\alpha_k(T, \mu)\,
    \hat{G}_k\quad.
\end{eqnarray}
Then the Ginzburg-Landau equation,
$\frac{\delta \Psi}{\delta \Delta^*(x)}=0$, to this order is precisely an
element of the AKNS hierarchy:
\begin{eqnarray}
    \sum_{k=0}^{N+1}\alpha_k(T, \mu)\, \frac{(k-1)}{2}\, \hat{f}_{k-2} =0
\end{eqnarray}
with the coefficients expressed in terms of the real-valued
functions $\alpha_k(T, \mu)$. Recall that
$\hat{f}_{-2}=\hat{f}_{-1}=0$.

As already mentioned, the elements of the AKNS hierarchy have some
remarkable properties, and because of the relation between the BdG
resolvent and the AKNS hierarchy, these properties turn out to be
very useful in solving physical problems based on the BdG
Hamiltonian.

For example, suppose $\Delta$ satisfies the $(N+1)^{\rm th}$ order
differential equation ${\rm AKNS}_N$. It automatically follows
from the recursion relations (\ref{recursion}) that $\Delta$
satisfies {\it all} the higher equations ${\rm AKNS}_{N+k}$, for
$k=1, 2, \dots$, with a suitable choice of the coefficients $c_k$.
This means that there is an infinite number of conserved
quantities for any member of the AKNS hierarchy.
Physically, this has the important implication  that it
is possible to make a consistent {\it polynomial} ansatz for the Gorkov
resolvent in order to solve the Dikii-Eilenberger equation (\ref{dikii}), as was done in \cite{bd2}.
Thus, if we make the {\it polynomial} ansatz for $R(x; E)$:
\begin{eqnarray}
R(x; E)=\sum_{n=0}^N \beta_n(E)\, \begin{pmatrix}
    {\hat{g}_n(x) & \hat{f}_{n-1}(x)\cr
    \hat{f}_{n-1}^*(x) & \hat{g}_n(x)}
    \end{pmatrix}
    \label{poly}
    \end{eqnarray}
    then $R(x; E)$ satisfies the Dikii-Eilenberger equation (\ref{dikii}) provided
    $\Delta(x)$ satisfies the ${\rm AKNS}_{N-1}$ equation.
    Therefore, given a solution of ${\rm AKNS}_{N-1}$, the exact
    resolvent can be written in a simple polynomial form (\ref{poly}),
    with the coefficients $\beta_n(E)$ determined from simple algebraic relations.
    This fact was used in \cite{bd2} to solve the gap equation in the ${\rm NJL}_2$
    model, using $N=2$ in which case the condensate function $\Delta(x)$
    had to satisfy the ${\rm AKNS}_{1}$ equation, otherwise known as the (complex)
    nonlinear Schr\"odinger equation . In fact, this {\it polynomial} form
    of the resolvent of the BdG Hamiltonian generalizes an analogous property
    of Schr\"odinger Hamiltonians. For example, it was known already to
    Hermite that for the Lam\'e potential, $V(x)=N(N+1)\,\nu \, {\rm sn}^2(x)$,
    with $N$ integer, the resolvent can be written as a polynomial of
    degree $N$ in $V(x)$ \cite{belokolos}.

Next we  connect this discussion to the known finite gap solutions
of the AKNS hierarchy \cite{akns,gesztesy}. First, we discuss the
spectral properties of the explicit finite gap solutions, and then
explain their relation to  nonlinear quantum mechanical
supersymmetry. The remaining sections will discuss details of
explicit cases where $\Delta$ is trigonometric (non-periodic case)
or elliptic (periodic case).

\subsection{Finite gap solutions to the AKNS hierarchy}

When $\Delta$ satisfies an equation from the AKNS hierarchy, the spectrum of the system
(\ref{ham-bdg}) acquires a finite-gap nature \cite{gesztesy,gw,smirnov}. Suppose we have a
complex \emph{periodic} $\Delta$ that obeys the ${\rm AKNS}_N$
equation [this picture can be extended for the case of a
quasi-periodic $\Delta$; see Section IV below].  The spectrum of
(\ref{ham-bdg}) is composed then of $N$ valence bands, $[E_{2k-1},
E_{2k}]$, $k=1,\ldots,N$, and by two conduction bands,
$(-\infty,E_0]$ and $[E_{2N+1},\infty)$, separated by $N+1$ energy
gaps,
\begin{equation}
    \sigma(H)=(-\infty, E_{0}]\cup[E_{1},
    E_{2}]\cup...\cup[E_{2N-1},E_{2N}]\cup[E_{2N+1},\infty )\quad .
\label{spectrum}
\end{equation}
The $2N+2$ energies $E_k$, $k=0,...,2N+1$, correspond to singlet
band-edge states, which are  described  either by
periodic or by anti-periodic wave functions.
The states in the interior of the allowed bands are described by
Bloch-Floquet quasi-periodic functions, and any corrresponding
energy level is doubly degenerate, corresponding physically to left-
and right-going waves. In the infinite period limit, $\Delta$
transforms into a non-periodic function, $N$ valence bands shrink
and transform into $N$ bound states;  the quasi-periodic states of
the conduction bands reduce to scattering states with energies
continuously varying in the intervals $(-\infty, E_{0}]$ and
$[E_{2N+1},\infty )$, where all energy levels, except edge values,
are doubly degenerate.

\subsection{Nonlinear quantum mechanical supersymmetry of AKNS${}_N$}

A double degeneration of the spectrum, accompanied by a presence of
a finite number of singlet states, is a typical property of ${\cal
N}=2$ nonlinear quantum mechanical supersymmetry \cite{andrianov,
Bosusy, nonlinear-susyPT, nonlinear-susyLame}. In order to identify
its structure for the finite-gap Bogoliubov-de Gennes system, we
first describe briefly different realizations of ${\cal N}=2$
supersymmetry in Schr\"odinger systems with a second order
Hamiltonian.

Quantum mechanical supersymmetry in its simplest form \cite{witten, reviewsusy,grant}
reflects coherently spectral properties of a  composed system,  given by
superpartner  Schr\"odinger
 Hamiltonians
 $\mathcal{H^{\pm}}=-d^2/dx^2+V^{\pm}(x)$,
$V^{\pm}(x)=W^2(x)\pm W'(x)$. The latter obey  intertwining
relations of a Darboux transformation \cite{matveev},
\begin{equation}
    A_1\mathcal{H}^-=\mathcal{H}^+A_1\quad, \qquad
    A_1^{\dagger}\mathcal{H}^+=\mathcal{H}^-A_1^{\dagger}\quad.
     \label{intert}
\end{equation}
Here $A_1=\psi_0 \frac{d}{dx} \frac{1}{\psi_0}=\frac{d}{dx}+W(x)$ is
a first order differential operator, whose kernel is a (generically,
formal) zero-value eigenstate $\psi_0(x)$ of the second order
differential operator ${\cal H^-}$. If $\psi_0(x)$ is a nodeless
function, the superpotential $W(x)=-\frac{d}{dx}\ln\psi_0(x)$ and both
superpartner potentials $V^{\pm}(x)$ are regular. Factorizations
${\cal H^-}=A_1^\dagger A_1$ and ${\cal H^+}=A_1 A_1^\dagger$
reflect a non-negative nature of the spectra of  both subsystems,
which in a generic case are almost isospectral by virtue of
intertwining relations (\ref{intert}). In non-periodic case, when
neither of the states $\psi_0$ and $1/\psi_0$ (the latter being a
kernel of the operator $A_1^\dagger$) is physical, i.e. is not a
normalizable function, two subsystems have exactly the same positive
definite spectrum. If one of the functions $\psi_0$ or $1/\psi_0$ is
normalizable, it is a singlet zero-energy state of the
supersymmetric system $\mathcal{H}=diag
(\mathcal{H^+},\mathcal{H}^-)$. In  periodic case both states
$\psi_0$ and $1/\psi_0$ are physical when they are  periodic, and
the lowest zero-energy level is doubly degenerate
\cite{SUSYperiodic}.

Intertwining and factorization relations are rewritten equivalently
in the form of a linear (Lie) ${\cal N}=2$ superalgebra
\begin{equation}\label{Liesusy}
    [\mathcal{Q}_a,\mathcal{H}]=0\quad,\qquad
    \{\mathcal{Q}_a,\mathcal{Q}_b\}=2\delta_{ab}\mathcal{H}\quad,
\end{equation}
generated by the matrix Hamitlonian ${\cal H}$ and supercharges
\begin{equation}\label{Qa}
    \mathcal{Q}_1=i\begin{pmatrix} {0&A_1\cr - A_1^{\dagger} & 0}
    \end{pmatrix}\quad,\qquad \mathcal{Q}_2=i\sigma_3 \mathcal{Q}_1\quad.
\end{equation}
The integral of motion $\Gamma=\sigma_3$, with $\Gamma^2=1$, plays
here the role of a $\Z_2$-grading operator,
\begin{equation}\label{Z2}
    [\Gamma,{\cal H}]=0\quad,\qquad
    \{\Gamma,Q_a\}=0\quad.
\end{equation}

Eq. (\ref{intert}) is generalized for the case of a Crum-Darboux
transformation \cite{matveev} generated by an intertwining operator
$A_n$ of order $n>1$, which annihilates $n$ linear independent
(generically, formal) eigenstates of ${\cal H^-}$. The partner
Hamiltonians are related by $\mathcal{H}^+=\mathcal{H}^--2(\ln {\cal
W}_n(\psi_1,...,\psi_n))''$, where ${\cal W}_n$ is the Wronskian of
the states of the kernel of $A_n$. If ${\cal W}_n(x)$ is a nodeless
function, the partner Hamiltonians ${\mathcal H}^\pm$ have the same
regularity properties. Intertwining relations are reformulated,
again, in the form of zero commutators of the supercharges
$\mathcal{Q}_a$, constructed in the form (\ref{Qa}) with $A_1$
substituted by $A_n$, with matrix Hamiltonian ${\cal H}={\rm
diag}({\mathcal H}^+, {\mathcal H}^-)$. In correspondence with
Burchnall-Chaundy theorem \cite{BCt, Ince}, the linear
superalgebraic relation (\ref{Liesusy}) becomes a nonlinear one,
\begin{equation}
     \{\mathcal{Q}_a,\mathcal{Q}_b\}=
     2\delta_{ab}P_n(\mathcal{H})\quad,
     \label{nonlinearsusy}
\end{equation}
where $P_n(\mathcal{H})$ is a polynomial of order $n$ in the
Hamiltonian.
For such a \emph{non-linear} supersymmetry,  of order $n$, the
operator $\Gamma=\sigma_3$ is again a $\Z_2$-grading operator. The
spectra of the two superpartner Schr\"odinger Hamiltonians
${\mathcal H}^\pm$ may differ in $0\leq l\leq n$ energy levels.

Some \emph{non-extended} systems with second order Hamiltonian are
characterized by a hidden supersymmetry of a bosonized nature
\cite{Bosusy, nonlinear-susyPT, nonlinear-susyLame}. In this case
${\cal H^+}={\cal H}^-$, and the intertwining relation reduces to
the commutativity of ${\cal H^-}$ with a Hermitian differential
operator which itself is identified as one of the supercharges,
${\cal Q}_1={\cal Z}$. Another supercharge is ${\cal Q}_2=i\Gamma
{\cal Q}_1$, where a $\Z_2$-grading operator $\Gamma$ is now a
parity (reflection), or a twisted parity
operator~\cite{nonlinear-susyPT}. A bosonized supersymmetry of a
nonlinear form appears, in particular, in finite-gap periodic and
reflectionless non-periodic second order systems, in which a
nontrivial operator of a Lax pair is identified as supercharge $Q_1$
annihilating all the singlet energy eigenstates.

Corresponding systems extended by means of a Darboux or a
Crum-Darboux transformation are described by a more rich,
tri-supersymmetric structure \cite{tri-delta, trisusy, trisusy1} . The latter
admits three alternative choices for $\Z_2$-grading operator,
$\Gamma=\sigma_3,\, {\cal R},$ or ${\cal R}\sigma_3$, where ${\cal
R}$ is a parity operator, and will be discussed in
more detail in the next sections.

For any type of  supersymmetry described above, infinitesimal
supersymmetric transformations generated by a supercharge ${\cal Q}$
can be presented in a  form
\begin{equation}\label{SUSYtrans}
    \delta {\cal X}=[\delta \alpha {\cal Q},{\cal X}]\quad,\qquad
    \delta\alpha=\xi\, \Gamma\quad,
\end{equation}
where  $\delta\alpha$ is a composition of a  Grassmann parameter
$\xi$ and of a $\Z_2$-grading operator $\Gamma$.

For the Bogoliubov-de Gennes system, Hamiltonian (\ref{ham-bdg}) is
a $2\times 2$ matrix first order differential operator. When (in a
generic case complex) the function $\Delta$ satisfies the ${\rm
AKNS}_N$ equation, the system displays a \emph{non-linear}
supersymmetry of order $(N+1)$. To identify its structure, we
consider the  $2\times 2$ matrix operator $S_{N+1}$ in
(\ref{q}) as one of the supercharges, which annihilates the whole
set of $(2N+2)$ band edge states,
  \begin{equation}
   S_{N+1}\psi_k=0\quad,
   \quad H\psi_k=E_k\psi_k\quad , \quad k=0,1...,2n+1\quad .
\label{anni}
\end{equation}
Using a matrix analogue of the Burchnall-Chaundy theorem, one
finds that the square of  $S_{N+1}$   produces a spectral
polynomial of degree $(2N+2)$
\footnote{For details and proof see Ref. \cite{gesztesy}},
\begin{equation}
    S^2_{N+1}=P_{2N+2}(H)=\prod_{k=0}^{2N+1}(H-E_k)\quad,
\end{equation}
associated with the hyperelliptic curve of genus $N$. We will show
that for any particular choice of the function $\Delta$ satisfying
the ${\rm AKNS}_N$ equation, there exists an integral $\Gamma$,
unitary equivalent to a \emph{nonlocal} operator ${\cal
R}\sigma_3$, with the properties
\begin{equation}
    [H,\Gamma]=0\quad , \qquad    \{S_{N+1},\Gamma\}=0 \quad,
    \qquad \Gamma^2=1\quad .
    \label{HbdGGamma}
\end{equation}
Identifying integral $\Gamma$ as a $\Z_2$-grading operator, we
treat  $S_{N+1}$  as one of the supercharges, ${\cal
Q}_1=S_{N+1}$, and define another supercharge in the usual way,
${\cal Q}_2=i\Gamma {\cal Q}_1$. As a result we find that the
system is described by ${\cal N}=2$ nonlinear supersymmetry,
\begin{equation}
    [\mathcal{Q}_a,H]=0\quad, \qquad
    \{\mathcal{Q}_a,\mathcal{Q}_b\}=2\delta_{ab}P_{2N+2}(H)\quad .
    \label{polybdg}
\end{equation}
The supersymmetric transformations are given by Eq.
(\ref{SUSYtrans}).

Notice that the described general structure of nonlinear
supersymmetry of the finite-gap Bogoliubov-de Gennes system is
similar to the hidden bosonized supersymmetry of a finite-gap
Schr\"odinger system \cite{nonlinear-susyPT, nonlinear-susyLame}. In
both cases nontrivial integrals annihilate all the singlet states,
all the other physical states are organized in energy doublets, and
the corresponding $\Z_2$-grading operators are nonlocal.

\section{Real defocusing hierarchy nS$_+$ and tri-supersymmetry}

In this Section we consider the real case $\Delta=\Delta^*$ of the
BdG system (\ref{ham-bdg}). As it follows from a unitary
transformation corresponding to a global chiral rotation with
constant parameter $\varphi$
\begin{equation}\label{chiralrot}
     U(\varphi) H(\Delta)U^{-1}(\varphi)=H(\tilde{\Delta})\quad,\qquad
     U(\varphi)=e^{\frac{i}{2}\varphi\gamma^5}\quad,\qquad \tilde{\Delta}=
    e^{i\varphi}\Delta\quad,
\end{equation}
a real case of $\Delta$  can be understood  modulo $U(1)$.

When $\Delta=\Delta^*$, the Hamiltonian anticommutes with
$i\gamma^1=\sigma_2$. As a consequence, the spectrum is symmetric
with respect to the energy reflection $E\rightarrow -E$. Then the
energies of singlet band-edge states (\ref{spectrum}) obey a
symmetry
\begin{equation}
    E_{2N+1-i}=-E_i\quad, \quad i=0,..,N\quad.
    \label{cond-spectrum}
\end{equation}

For the following discussion it is convenient to present Hamiltonian
(\ref{ham-bdg}) in another form by applying a unitary transformation
generated by a matrix
\begin{equation}
    \mathcal{U}=\frac{1}{2}\begin{pmatrix} {1+i & 1-i\cr \ -1-i & 1-i}
    \quad .
\end{pmatrix}
\label{unitary1}
\end{equation}
This amounts to a  change of Dirac representation from
(\ref{gamma-1}) to
\begin{eqnarray}
    \gamma^0=-\sigma_2
    \quad , \quad
    \gamma^1=-i\sigma_3
    \quad , \quad
    \gamma^5=-\sigma_1 \quad.
\label{gamma-22}
\end{eqnarray}
The BdG Hamiltonian (\ref{ham}) with real $\Delta$ takes therefore a
form
\begin{equation}
    H_{\R}=i\begin{pmatrix}
    {0 & \frac{d}{dx}+\Delta(x)\cr \ \frac{d}{dx}-\Delta(x)& 0}
    \quad .
\end{pmatrix}
\label{realH}
\end{equation}
This coincides with one of the matrix supercharges (\ref{Qa}) of a
supersymmetric quantum mechanicanical system \cite{witten} described
by a matrix Schr\"{o}dinger Hamiltonian $\mathcal{H}=H_\R^2$, for
which $\Delta=W(x)$ is identified as the superpotential,  and
supersymmetric partner potentials are given by
\begin{equation}
    V^{\pm}(x)=\Delta^2 \pm \Delta' \quad.
\label{deltasuper}
\end{equation}
In accordance with (\ref{gamma-1}) [and also (\ref{gamma-2})
below], the operator $i\gamma^1$ underlying the energy reflection
symmetry of the real BdG system corresponds to the grading
operator $\Gamma=\sigma_3$ for the associated superextended
Schr\"odinger system.

If $\Delta$ satisfies the {\rm AKNS}${}_N$ equation, the number of
gaps modulo $2$ in the spectrum of the  periodic BdG system (or, of
bound states in the non-periodic limit) depends on the properties of
periodicity (normalizability) of the functions
\begin{equation}
    \Psi^{\pm}_0(x)=e^{\pm \int^x_0 \Delta(y) dy}\, u^\pm \quad,
\label{psi-delta}
\end{equation}
where $u^+=(1,0)^t$, $u^-=(0,1)^t$, and $t$ means  transposition.
 Namely, if $N=2m-1$, $m \in \mathbb{N}$, and both functions
(\ref{psi-delta})  are periodic (or, one of these functions is
normalizable), they describe two (or, one in a non-periodic case)
zero energy physical states of the system (\ref{realH}) with
\emph{even} number of gaps in a spectrum. For $N=2m$, $m\in\N$,
states (\ref{psi-delta}) are not periodic (normalizable in the
non-periodic limit), and level $E=0$ does not belong to the spectrum
characterized by \emph{odd} number of gaps.

{}From the viewpoint of the associated super-extended
Schr\"{o}dinger Hamiltonian
\begin{equation}
    \mathcal{H}=H_\R^2\quad ,
    \label{HHR}
\end{equation}
the cases $N=2m-1$ and $N=2m$ correspond to finite-gap systems
with exact and broken ${\cal N}=2$ supersymmetries, respectively.

When real $\Delta$ satisfies an AKNS equation, the equality
(\ref{HHR}), that connects the BdG and extended Schr\"{o}dinger
Hamiltonians, reflects an intimate relation between the AKNS and
Korteweg- de Vries (KdV) hierarchies (see for example
\cite{gesztesy}). This is just because in this case the
Schr\"odinger potentials $V^{\pm}$ satisfy equations of the
corresponding KdV stationary hierarchies. Therefore, each component
${\cal H^{\pm}}$ of the diagonal Schr\"{o}dinger Hamiltonian ${\cal
H}$ has a nontrivial integral of motion in the form of a
differential operator ${\cal Z^{\pm}}$ from the corresponding Lax
pair $({\cal H^{\pm}},{\cal Z^{\pm}})$ of the KdV hierarchy. The
diagonal matrix operator ${\cal Z}=diag({\cal Z}^{+},{\cal Z}^{-})$
commutes with the matrix Hamiltonian, $[{\cal H},{\cal Z}]=0$, and
leads to existence of a more rich, tri-supersymmetric structure
\cite{trisusy, trisusy1},  which will be discussed in more detail
below. The systems with  exact and broken ${\cal N}=2$ supersymmetry
reveal the tri-supersymmetric structure in different ways, and these
two cases will be considered separately.

\subsection{ Exact ${\cal N}=2$ supersymmetry and tri-supersymmetry in the light of the nS$_+$
hierarchy}\label{IIIA}

Let us start with the case of unbroken ${\cal N}=2$ supersymmetry,
and consider the BdG system  (\ref{realH}) given by a real
\emph{odd} function $\Delta(x)$,
\begin{equation}
    \Delta(-x)=-\Delta(x)\quad ,
\label{Dodd}
\end{equation}
satisfying the $(2m)^{\rm th}$ order equation ${\rm AKNS}_{2m-1}$,
$m\in \N$. Since Hamiltonian $H_{\R}$ anticommutes with both
$\sigma_3$ and ${\cal R}$,  the operator ${\cal R}\sigma_3$ is a
non-local integral of motion, $[H_{\R},{\cal R} \sigma_3]=0$. The
stationary $2m-1$-th order AKNS hierarchy is described by the Lax
pair $(H_\R, S_{2m})$, where  $S_{2m}$ is a non-trivial integral of
motion, $[H_\R, S_{2m}]=0$. In the present case it anticommutes with
the trivial integral ${\cal R} \sigma_3$, $\{S_{2m}, {\cal R}
\sigma_3\}=0$. We can treat  ${\cal R} \sigma_3$ as the grading
operator, and $S_{2m}$ as a non-linear supercharge for the BdG
Hamiltonian in accordance with relations (\ref{polybdg}) and
(\ref{HbdGGamma}), where $N=2m-1$.

Because of the presence of the nontrivial integral $S_{2m}$ in the
BdG system, the associated extended Schr\"odinger system with
parity-even Hamiltonian ${\cal H}$ is characterized by a more rich
than ${\cal N}=2$ supersymmetry structure. Indeed,  relation
(\ref{HHR}) implies that the matrix operator $S_{2m}$ is also the
integral of motion for the associated Schr\"odinger system. Denoting
$H_{\R}={\cal X}$ and $S_{2m}={\cal Y}$, we have
\begin{equation}
    [{\cal H},{\cal X}]=[{\cal H},{\cal Y}]=0\quad ,\qquad
    [{\cal X},{\cal Y}]=0\quad,
    \label{triXY}
\end{equation}
i.e. ${\cal X}$ and ${\cal Y}$ are commuting parity-odd and
parity-even integrals, respectively.
 As we noted above, the extended Schr\"odinger system ${\cal
H}$ is characterized also by the diagonal higher order matrix
integral ${\cal Z}$. Its relation with the original BdG system can
easily be established. Consider the next after $S_{2m}$ integral
$S_{2m+1}$ from the infinite set of integrals of the corresponding
AKNS hierarchy. For the BdG Hamiltonian $H_{\R}$, operator
$S_{2m+1}$ appears as a trivial composite integral,
\begin{equation}
    S_{2m+1}=H_{\R}S_{2m}=S_{2m}H_{\R}\quad.
    \label{BdGtri}
\end{equation}
For the associated  extended Schr\"odinger system, this is exactly
the nontrivial diagonal parity-odd integral, $S_{2m+1}={\cal Z}$,
\begin{equation}
    {\cal Z}={\cal X}{\cal Y}={\cal Y}{\cal X}\quad .
    \label{ZXYex}
\end{equation}
Then relations (\ref{triXY}) can be
supplied by
\begin{equation}
    [{\cal H},{\cal Z}]=0\quad,\qquad [{\cal Z},{\cal X}]=[{\cal Z},{\cal
    Y}]=0\quad .
\end{equation}
Operators ${\cal X}$, ${\cal Y}$ and ${\cal Z}$ constitute a part of
nontrivial  integrals for the Schr\"odinger system. Hamiltonian
${\cal H}$ commutes also with mutually commuting trivial integrals
$\sigma_3$, ${\cal R}$ and ${\cal R}\sigma_3$, and their products
with ${\cal X}$, ${\cal Y}$ and ${\cal Z}$ produce other nontrivial
integrals of motion. All these operators together generate the
tri-supersymmetry of the second order finite-gap system ${\cal H}$.
The resulting complete nonlinear superalgebraic structure depends on
the identification of one of the trivial integrals, $\sigma_3$,
${\cal R}$, or ${\cal R}\sigma_3$, as the grading operator. For any
of the three possible choices, exactly one of the nontrivial
integrals ${\cal X}$, ${\cal Y}$, or ${\cal Z}$ is identified as
$\Z_2$-even operator, which plays the role of the central charge of
the superalgebra, while other two are treated as $\Z_2$-odd
supercharges. This is in contrast with the supersymmetric structure
of the BdG system $H_{\R}$, for which the only unitary operator
${\cal R}\sigma_3$ can be chosen as the grading operator. For the
details on the tri-supersymmetry we refer the reader to
\cite{trisusy,trisusy1}.

Next we discuss some concrete examples of solutions and the associated explicit form
of the nonlinear supersymmetric structure for the real ${\rm AKNS}_N$ systems with  $N=2m-1$, $m\in\N$.

\subsubsection{Real periodic solution of the $(2m-1)$th nS$_{+}$ equation}
\begin{figure}[h!]
\centering
\includegraphics[scale=0.835]{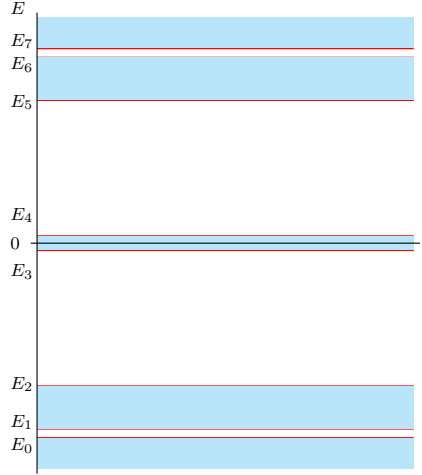}
\caption{Spectrum of (\ref{uniale}) for $m=2$ and with $\nu=0.9$.
Red lines are the singlet band edge energies. The black line
represents the zero energy level, with respect to which all the
spectrum is symmetric.} \label{FigExac}
\end{figure}

Consider a periodic solution of the $(2m-1)^{\rm th}$ nS$_{+}$
equation \footnote{For its applications see \cite{bd2, bdt, DGKL}
and references therein.},
\begin{equation}
    \Delta = m \, \nu \, \frac{{\rm sn}\,
    (x, \nu) \,{\rm cn}\, (x, \nu) }{{\rm dn}\, (x, \nu)}\quad,
\label{nsn}
\end{equation}
where $m \in \mathbb{N}$, and ${\rm sn}\, (x, \nu) ={\rm sn}\,x$,
${\rm cn}\, (x, \nu)={\rm cn}\, x$  and  ${\rm dn}\, (x, \nu)={\rm
dn}\, x$ are the Jacobi elliptic functions of elliptic
parameter $\nu$, with $0< \nu< 1$ \cite{lawden}.
 The functions ${\rm cn}\, x$ and ${\rm dn}\, x$
are even, while ${\rm sn}\, x$ is odd, and therefore (\ref{nsn}) is
an odd function.  In this case the Hamiltonian (\ref{realH}) takes
the form of an odd operator
\begin{equation}
    H_\R=i\begin{pmatrix}
    {0 & \frac{d}{dx}+ m\, \nu \frac{{\rm sn}\,x\,{\rm cn}\,x }
    {{\rm dn}\, x}    \cr \ \frac{d}{dx}-m \,\nu \frac{{\rm sn}\,x\,{\rm cn}\,x
    }{{\rm dn}\, x}  & 0}\quad.
\end{pmatrix}
\label{uniale}
\end{equation}
Therefore, the system is characterized  by a nonlocal integral of
motion
\begin{equation}\label{GammaR}
    \Gamma={\cal R}\sigma_3\quad.
\end{equation}
Hamiltonian (\ref{uniale}) has a real period $T=2 {\bf K}(\nu)$,
$H_\R(x)=H_\R(x+T)$, where ${\bf K}(\nu)$ is the complete elliptic
integral of the first kind. This system has $4m$ band edge-states of
non-degenerated energies $E_i$, $i=0,...,4m-1$, distributed
symmetrically  [see Figure \ref{FigExac} ],
\begin{equation}\label{E4sym}
    E_{4m-1-i}=-E_i\quad,
    \qquad i=0,..,2m\quad,
\end{equation}
with respect to the doubly degenerate zero energy level, which is in
the middle of the central allowed band. All the band-edge singlet
states are anti-periodic, their period is $4{\bf K}(\nu).$

With $\Delta$ satisfying  the $(2m-1)^{\rm th}$ nS$_{+}$ equation,
the Hamiltonian (\ref{uniale}) commutes with a non-trivial integral
of motion (\ref{q}), which can be presented in a form
\begin{equation}
    S_{2m}=\begin{pmatrix}
    {0 & S^-_{2m}\cr \  S^+_{2m}  & 0}
\end{pmatrix}\quad,
\label{qperiodic}
\end{equation}
where
\begin{equation}
    S^-_{2m}=\frac{{{\rm dn}^{m+1} x}}{{\rm cn}^{2m+1} x}
    \left(\frac{{\rm cn}^{2} x}{{\rm dn}\, x}\frac{d}{dx}\right)^{2m}
    \frac{{\rm dn}^{m-1} x}{{\rm cn}^{2m-1} x}\quad, \qquad
    S^+_{2m}=(S^-_{2m})^{\dagger}\quad.
\end{equation}
Since $S^-_{2m}$ is an \emph{even} differential operator,
(\ref{qperiodic})  anti-commutes with the integral (\ref{GammaR}),
$\{S_{2m},\mathcal{R}\sigma_3\}=0$. Operator (\ref{GammaR}) is
identified as a $\Z_2$-grading operator, and integrals
\begin{equation}\label{Q12per}
    \mathcal{Q}_1=S_{2m}\quad,
    \qquad \mathcal{Q}_2=i\mathcal{R}\sigma_3\mathcal{Q}_1
\end{equation}
play the role of the supercharges for the system (\ref{uniale}).
They generate the ${\cal N}=2$ polynomial superalgebra
\begin{eqnarray}
    [\mathcal{Q}_a,H_\R]=0\quad , \qquad \{\mathcal{Q}_a,\mathcal{Q}_b\}=
    2\delta_{ab}P_{4m}(H_\R)\quad,&
    \label{superalgebra}
\end{eqnarray}
where, in correspondence with (\ref{E4sym}),
\begin{eqnarray}\label{nsusy4m}
     &P_{4m}(H_\R) =\prod_{i=0}^{4m-1}
     (H_\R-E_i)=\prod_{i=0}^{2m-1}(H_\R^2-E_i^2)\quad.&
\end{eqnarray}
The associated with (\ref{uniale})  superextended Schr\"odinger
system given by the second order Hamiltonian ${\cal H}=H_\R^2$
describes, in correspondence with Eq. (\ref{E4sym}),  the $m$-gap
self-isospectral Lam\'e system \cite{SUSYperiodic, trisusy,
trisusy1} with band-edge energies $\tilde{E}_0=0$,
$\tilde{E}_j=E_{2m-1+j}^2=E_{2m-j}^2>0$, $j=1,\ldots,2m$.

For the BdG Hamiltonian in the simplest case $m=1$, the singlet
band-edge states and their energies are given by
\begin{eqnarray}
    &\psi_{0}=\begin{pmatrix}
    { i {\rm cn}\,(x+b/2) \cr  -{\rm cn}\, (x-b/2)}
    \end{pmatrix}\quad, \quad \psi_{1}=\begin{pmatrix}
    { i {\rm sn}\,(x+b/2) \cr  -{\rm sn}\, (x-b/2)}
    \end{pmatrix}\quad, \quad \psi_2=\sigma_3\psi_1\quad , \quad
    \psi_3=\sigma_3\psi_0\quad,&
    \\
    &H_\R\psi_i=E_i\psi_i\quad,\quad
    E_{0}=-1\quad ,
    \quad E_{1}=-\sqrt{1-\nu}\quad, \quad E_{2}=- E_{1}\quad,
    \quad   E_{3}=- E_{0}\quad.& \label{bandener}
\end{eqnarray}
The explicit form of the states in the interior of the allowed bands
can be found, for example, in refs. \cite{whittaker}.

\subsubsection{Real non-periodic solution of the $(2m-1)$th nS$_{+}$ equation}

Taking the infinite period limit of (\ref{nsn}), which corresponds
to $\nu \rightarrow 1$, we get  a non-periodic solution of the
$(2m-1)$th  nS$_{+}$ equation,  the familiar kink
\begin{equation}
    \Delta= m \tanh x \quad, \quad m \in \mathbb{N} \quad.
\label{npt}
\end{equation}
Hamiltonian (\ref{realH}) takes a form
\begin{equation}
    H_\R=i\begin{pmatrix}
    {0 & \frac{d}{dx}+m\tanh x\cr \ \frac{d}{dx}-m\tanh x& 0}
    \end{pmatrix}=i\begin{pmatrix}
    {0  & \mathcal{D}_m \cr \mathcal{D}_{-m} & 0}
    \end{pmatrix}\quad,
\label{unipt}
\end{equation}
where the operator $\mathcal{D}_{\lambda}$ satisfies the relations
$ \mathcal{D}_{-\lambda}=-\mathcal{D}^{\dagger}_{\lambda}$ and
$\mathcal{D}_{-\lambda}\mathcal{D}_{\lambda}=
\mathcal{D}_{\lambda+1}\mathcal{D}_{-\lambda-1}+2\lambda+1$.

The central band $[E_{2m-1},E_{2m}]$, $E_{2m-1}=-E_{2m}$, of the
periodic system  shrinks here into a zero energy level,
$[E_{2m-1},E_{2m}] \rightarrow \mathcal{E}_m=0$, and the
corresponding band-edge states transform into a  single bound state.
The other valence bands shrink as well, producing $(2m-1)$ bound
states of non-zero values of energy: $[E_{1},E_{2}] \rightarrow
\mathcal{E}_1$, $[E_{3},E_{4}] \rightarrow \mathcal{E}_2$, etc. The
conduction bands transform into scattering bands with singlet edge
energies $\mathcal{E}_{0}<0$ and
$\mathcal{E}_{2m}=-\mathcal{E}_{0}>0$, where $E_{0}\rightarrow
\mathcal{E}_{0}$, $E_{4m-1}=-E_0 \rightarrow \mathcal{E}_{2m}$.

Defining the higher order differential operators
\begin{equation}\label{psisin}
    \mathcal{D}_{-\lambda}^\beta=\mathcal{D}_{-\lambda}\mathcal{D}_{-\lambda+1}...
    \mathcal{D}_{-\lambda+\beta-1}\quad, \quad \beta=2,...,\lambda \quad,
    \quad
    \mathcal{D}_{-\lambda}^0=1, \quad
    \mathcal{D}_{-\lambda}^1=\mathcal{D}_{-\lambda}\quad,
\end{equation}
we can write easily the eigenfunctions of (\ref{unipt}). The $2m+1$
singlet states acquire a form
\begin{eqnarray}\label{eigenRB}
    \psi_{m,l}^{\pm}(x)=\begin{pmatrix}
    {\mp\, i\, \mathcal{E}_{m-l}\, \mathcal{D}_{-m+1}^{l-1} \,
    {\rm sech}^{m-l}\,x \cr \mathcal{D}_{-m}^l \, {\rm sech}^{m-l}\,x  }
    \end{pmatrix}\quad ,
    \quad l=0,...,m\quad,
    \\  \quad H_\R \,
    \psi_{m,l}^{\pm}(x)=\pm \, \mathcal{E}_{m-l}
    \psi_{m,l}^{\pm}(x) \quad ,\quad
    \mathcal{E}_{m-l}=\sqrt{m^2-(m-l)^2}\quad.
\end{eqnarray}
Here $l=m$ corresponds to two singlet edge states of the two
symmetric parts of the continuous spectrum, while $l=0,..,m-1$ give
$(2m-1)$ bound states. For $l=0$, wave functions $\psi_{m,0}^{+}(x)$
and $\psi_{m,0}^{-}(x)$ coincide,
$\psi_{m,0}^{+}(x)=\psi_{m,0}^{-}(x)=\psi_{m,0}(x)$,
\begin{equation}
    \psi_{m,0}(x)=\begin{pmatrix} {0 \cr {\rm sech}^m\,x }
    \end{pmatrix}\quad.
    \label{psim0}
\end{equation}
This function describes a singlet bound state with energy
$\mathcal{E}_m=0$ in the middle of the spectrum.
 The scattering states of the bands $(-\infty, \mathcal{E}_0]$ and
$[\mathcal{E}_{2m},\infty)$ have a form
\begin{eqnarray}\label{eigenRS}
    \psi_{m; \uparrow (\downarrow)}^{\pm}(x)&=&\begin{pmatrix}
    {-(+)\, i\, \mathcal{E}_{m;k}\, \mathcal{D}_{-m+1}^{m-1} \,
    e^{\pm ikx} \cr \mathcal{D}_{-m}^m \, e^{\pm ikx}  }
    \end{pmatrix}\quad, \\
    H_\R \,\psi_{m;\uparrow}^{\pm}(x)= \, \mathcal{E}_{m;k} \psi_{m;
    \uparrow}^{\pm}(x)\quad, \quad H_\R \,\psi_{m;\downarrow}^{\pm}(x)&=& \,
    -\mathcal{E}_{m;k} \psi_{m; \downarrow}^{\pm}(x) \quad , \quad
    \mathcal{E}_{m:k}=\sqrt{m^2+k^2}\quad,
\end{eqnarray}
where $k> 0$. Boundary value $k=0$ reproduces singlet edge states
(\ref{psisin}) with  $l=m$.

The non-trivial integral (\ref{qperiodic}) transforms into
\begin{equation}
    S_{2m}=\begin{pmatrix}
    {0 & ( \mathcal{D}_{-m}^{2m} )^{\dagger} \cr \  \mathcal{D}_{-m}^{2m}  & 0}
    \end{pmatrix}=\begin{pmatrix}
    {0 & \mathcal{D}_{-m+1}\mathcal{D}_{-m+2}...\mathcal{D}_{m-1}\mathcal{D}_{m}
    \cr \  \mathcal{D}_{-m}\mathcal{D}_{-m+1}...\mathcal{D}_{m-2}\mathcal{D}_{m-1}  & 0}
    \end{pmatrix}.
\label{Qnon}
\end{equation}
The supercharges are identified as in (\ref{Q12per}). They
annihilate all the $2m+1$ singlet states. The nonlinear superalgebra
is given by the same equation (\ref{superalgebra}) as in the
periodic case, but with the spectral polynomial operator changed {
coherently with the bands shrunk,
\begin{equation}
    P_{4m}(H_\R)=(H_\R^2-\mathcal{E}_{0}^2)(H_\R-
    \mathcal{E}_{m})^2\prod_{i=1}^{m-1}(H_\R^2-\mathcal{E}_{i}^2)^2=(H_\R-\mathcal{E}_{0})
    (H_\R-\mathcal{E}_{2m})\prod_{i=1}^{2m-1}(H_\R-\mathcal{E}_{i})^2\quad.
\label{q2nonperiodic}
\end{equation}
The relation $S_{2m}^2=P_{4m}(H_\R)$ corresponds to a degenerate
hyperelliptic curve.

The associated second order supersymmetric Schr\"odinger system
describes two reflectionless P\"oschl-Teller systems with $m$ and
$(m-1)$ bound states in their spectra.  At $m=1$ one of the two
Schr\"odinger subsystems is a non-relativistic free particle
\footnote{Hidden nonlinear supersymmetry of the reflectionless
P\"oschl-Teller system can be explained in terms of the
Aharonov-Bohm effect for non-relativisitc particle on AdS(2), see
\cite{AdS2}.}.

\subsection{ Broken ${\cal N}=2$ supersymmetry and tri-supersymmetry
in the light of the nS$_+$ hierarchy}

Now we consider the BdG system (\ref{realH}) defined by real
$\Delta$ satisfying the AKNS${}_N$ equation with $N=2m$. As we
noted above, in this case the associated extended Schr\"odinger
system (\ref{HHR})  is characterized  by the broken ${\cal N}=2$
supersymmetry. Our discussion will be restricted by three-gap BdG
systems corresponding to the simplest case of $m=1$, for which the
explicit solution of the AKNS equation can be found in both
periodic and non-periodic cases. This  will provide us,
particularly, with a new structure of tri-supersymmetry, to which
the ${\cal N}=2$ supersymmetry of the  associated one-gap
Schr\"odinger system is extended. One of its peculiar features is
that in periodic and non-periodic cases the super-partner
potentials are given by the same function shifted for an
\emph{arbitrary} distance $b$.

In the examples we discuss below, the nonlinear supersymmetric
structure of the BdG system is based on the existence of the
non-trivial local integral $S_3$  and a trivial nonlocal integral
$\Gamma={\cal R}\sigma_2$ playing the role of  the $\Z_2$-grading
operator. The first integral is a consequence of the AKNS${}_2$
equation satisfied by $\Delta$, while the second means that it is
an even function, $\Delta(-x)=\Delta(x)$. Applying a chiral
rotation, the latter integral can be reduced to the form of the
grading operator we used in previous examples, i. e. ${\cal
R}\sigma_3$. We  do not apply, however, such a rotation here but
work with the BdG Hamiltonian having the fixed off-diagonal form
(\ref{realH}).

From the viewpoint of the associated extended Schr\"{o}dinger
system, the non-trivial  diagonal $2\times 2$  matrix integral
\begin{equation}
    {\cal Z}=S_3\quad,
\end{equation}
which is the third order differential operator, is related with the
stationary KdV hierarchy given by the Lax pair $(\mathcal{H},
\mathcal{Z})$, $[\mathcal{H}, \mathcal{Z}]=0$, where ${\cal H}={\cal
X}^2$, ${\cal X}=H_\R$. At the same time the next after $S_3$
trivial integral $S_4=S_3 H_\R=H_\R S_3$ for the BdG system  is also
a  trivial integral for the Schr\"odinger system, $S_4={\cal Z}{\cal
X}={\cal X}{\cal Z}$, cf. (\ref{BdGtri}) and (\ref{ZXYex}).

Let us construct  the matrix second order differential operator
$S_2$ according to (\ref{qnS}). A direct check shows that, as it
should be expected, this operator is not an integral of motion for
the BdG system,
\begin{equation}
    [H_\R, S_2]=i\epsilon (b) \sigma_3\quad,
\end{equation}
where $\epsilon (b)$ is some function of the displacement parameter,
see Eqs. (\ref{eps1}) and (\ref{eps2}) below. But since ${\cal
X}=H_\R$ anticommutes with $\sigma_3$ (remember that this operator
is a usual linear supercharge for the associated ${\cal N}=2$
superextended Schr\"odinger system ${\cal H}$, for which $\sigma_3$
is the grading operator), we get the relations
\begin{eqnarray}
    &[{\cal H},{\cal X}]=[{\cal H},{\cal Y}]=[{\cal H},{\cal
    Z}]=0\quad,&\label{tri1}\\
    &[{\cal Z}, {\cal X}]= [{\cal Z}, {\cal
    Y}]=0\quad,&\label{tri2}\\
    &[{\cal X}, {\cal Y}]=i\epsilon(b) \sigma_3\quad,&\label{tri3}
\end{eqnarray}
where ${\cal Y}=S_2$. Therefore, as in the case of the exact ${\cal
N}=2$ supersymmetry, the three operators ${\cal X}$, ${\cal Y}$ and
${\cal Z}$ are nontrivial integrals of motion for the associated
Schr\"odinger system.  With respect to the trivial integral
$\sigma_3$ identified as the $\Z_2$-grading operator $\Gamma$,
integrals ${\cal X}$ and  ${\cal Y}$ are fermionic supercharges of
the tri-supersymmetric structure, while ${\cal Z}$ is its bosonic
central charge. Unlike the case of the exact ${\cal N}=2$
supersymmetry discussed above, here the fermionic supercharges do
not commute, see (\ref{tri3}). Nevertheless, in both cases they
satisfy a relation
\begin{equation}\label{XYZa}
    \{{\cal X},{\cal Y}\}=2{\cal Z}\quad .
\end{equation}
Since the BdG Hamiltonian $H_\R$ anticommutes with $\sigma_3$ and
${\cal R}\sigma_1$,  these operators alongside with ${\cal
R}\sigma_2$ are the trivial integrals of motion  for the  associated
Schr\"odinger system ${\cal H}$. Therefore, besides the usual
identification $\Gamma=\sigma_3$, nonlocal operators ${\cal
R}\sigma_1$ and ${\cal R}\sigma_2$ can also be chosen as
$\Z_2$-grading operators for the  associated Schr\"odinger system.
Under the choice $\Gamma={\cal R}\sigma_1$, the operators ${\cal X}$
and ${\cal Z}$ are identified as fermionic supercharges, while
${\cal Y}$ plays the role of the bosonic integral for the system
${\cal H}$. For alternative choice $\Gamma={\cal R}\sigma_2$, ${\cal
Z}$ is again a fermionic supercharge, while the (boson-fermion)
nature of the operators ${\cal X}$ and ${\cal Y}$ is interchanged.
For these two choices of the grading operator the commutator
(\ref{tri3}) is a part of a complete list of (anti)-commutation
relations of the tri-supersymmetry of the Schr\"odinger system.

Let us stress that the essential  difference of the described
tri-supersymmetric structure of the Schr\"odinger system ${\cal
H}$ in comparison with the tri-supersymmetry  from Section
\ref{IIIA} is that here the three trivial unitary involutive
integrals of motion ${\cal R}\sigma_1$, ${\cal R}\sigma_2$ and
$\sigma_3$ anticommute mutually and satisfy between themselves the
same set of algebraic relations as the three Pauli sigma matrices.
At the abstract level, they generate non-abelian quaternion group
of order eight in contrast with the Klein four-group generated by
trivial integrals of the tri-supersymmetric structure of the
extended Schr\"odinger systems associated with the family of the
finite-gap BdG systems discussed above.

\subsubsection{ Real periodic solution of the second nS$_+$ equation}

Let us consider the periodic function
\begin{eqnarray}\label{realperiodic2}
    \Delta(x)&=&\frac{{\rm cn}\,b\, {\rm dn}\,b}{{\rm sn}\,b}+\nu\, {\rm sn}\,b\,{\rm sn}
    \,(x-b/2) \,{\rm sn}\,\left(x+b/2\right) \nonumber \\
    &=&Z(b;\nu)+\frac{{\rm cn}\,b\, {\rm dn}\,b}{{\rm
    sn}\,b}+Z(x-b/2;\nu)-Z(x+b/2;\nu)\quad,
\end{eqnarray}
where $Z(x;\nu)$ is the Jacobi Zeta function, and the real elliptic
parameter is chosen, $0<\nu < 1$. The real parameter $b$ lies in the
range $0\leq b\leq {\bf K}(\nu)$, where ${\bf K}(\nu)$ is the
elliptic half-period. The function (\ref{realperiodic2}) is a
solution of the third-order AKNS$_{2}$ equation (to be more
specific, in the terminology of \cite{gesztesy}, the second nS$_+$
equation):
\begin{equation}
    \Delta^{\prime \prime \prime}-6\Delta^2
    \Delta^{\prime}-2(1+\nu-3{\rm ns}^2\,b)\Delta'=0\quad.
\end{equation}
The Hamiltonian (\ref{realH}) takes the form
\begin{equation}\label{4bandperiodic}
    H_\R=i\begin{pmatrix} {0 & \frac{d}{dx}+ \frac{{\rm cn}\,b\, {\rm
    dn}\,b}{{\rm sn}\,b}+\nu\, {\rm sn}\,b\,{\rm sn}\,(x-b/2) \,{\rm
    sn}\,\left(x+b/2\right)   \cr \ \frac{d}{dx}-(\frac{{\rm cn}\,b\,
    {\rm dn}\,b}{{\rm sn}\,b}+\nu\, {\rm sn}\,b\,{\rm sn}\,(x-b/2)
    \,{\rm sn}\,\left(x+b/2\right))& 0}
    \end{pmatrix}\quad,
\end{equation}
and we can check that it has a trivial integral of motion
$\Gamma={\cal R}\sigma_2$, $[{\cal R}\sigma_2,H_\R]=0$,  where
${\cal R}$ means reflection. The square of (\ref{4bandperiodic})
produces a self-isospectral Schr\"{o}dinger operator with potentials
$V^\pm=\Delta^2\pm \Delta^\prime$ as in (\ref{deltasuper}),
satisfying the translation property $V^{-}(x)=V^{+}(x-b)$:
\begin{equation}\label{hisosper}
    H_R^2=\begin{pmatrix} {-\frac{d^2}{dx^2} +{\rm
    ns}^2\,b-1-\nu+2\nu\,{\rm sn}^2\,(x+b/2) & 0 \cr \  0 &
    -\frac{d^2}{dx^2}+{\rm ns}^2\,b-1-\nu+2\nu\,{\rm sn}^2\,(x-b/2)}
    \end{pmatrix}\quad .
\end{equation}

The spectrum of this system [see Figure \ref{FigBro}] is
composed of four bands, arranged symmetrically with respect to the
zero energy level, which does not belong to the physical spectrum.

\begin{figure}[h!]
\centering \label{FigBro1} \subfigure[$\nu=0.2, b=1$] {
\includegraphics[scale=0.9]{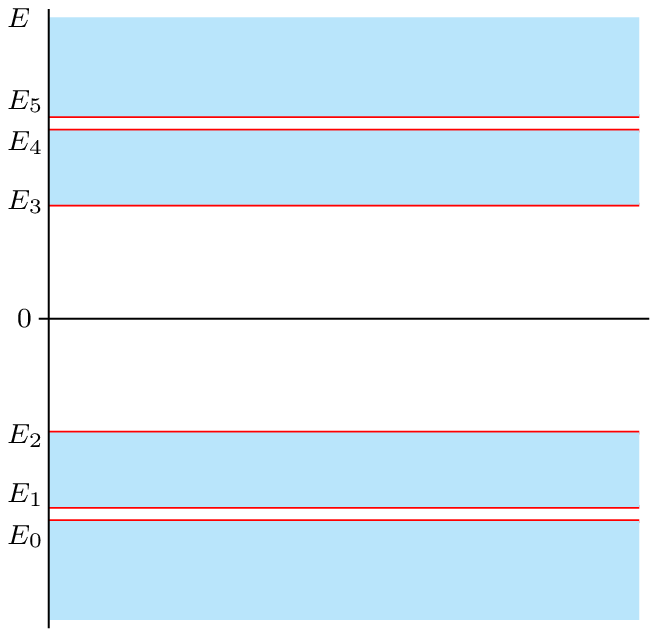}}
\label{FigBro2}\quad\quad \subfigure[$\nu=0.9, b=1$ ]
{\includegraphics[scale=0.9]{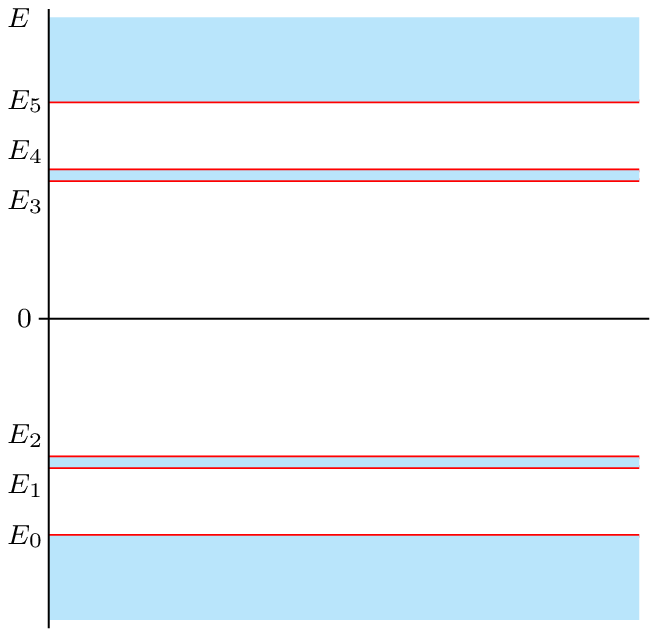}} \label{FigBro3}
\subfigure[$\nu=0.7, b=1$ ]
{\includegraphics[scale=0.9]{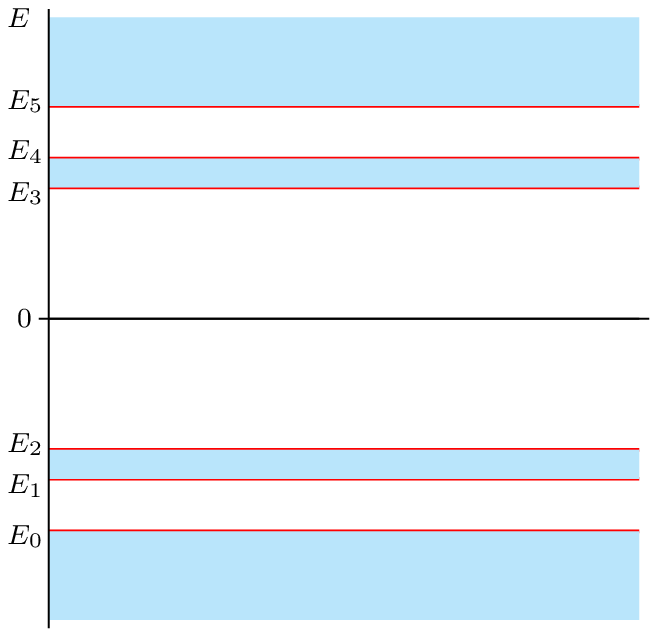}} \quad\quad
\label{FigBro4} \subfigure[$\nu=0.7, b={\bf K}(0.7)$ ]
{\includegraphics[scale=0.9]{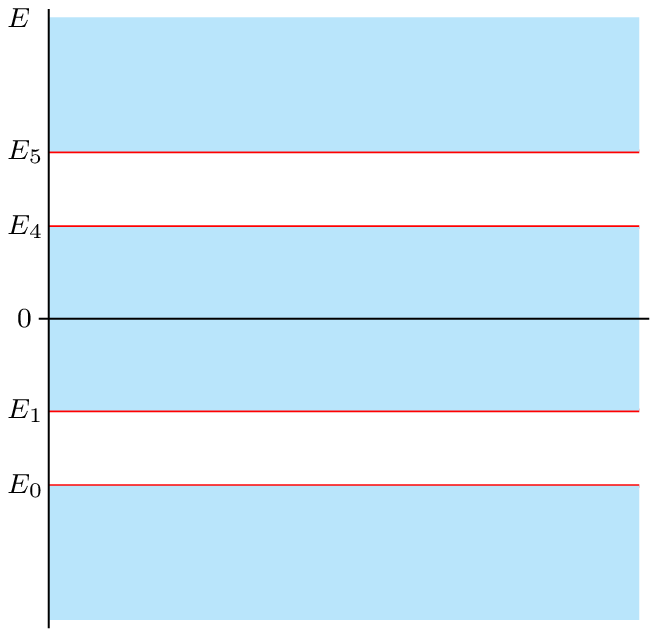}} \caption{Plots show the
spectrum of (\ref{4bandperiodic}) and its dependence on $\nu$ and
$b$. The red lines correspond to the band edge energies. Figures (a)
and (b) show the cases with different values of the modular
parameter $\nu$ and the same value of $b$; in figures (c) and (d)
the displacement constant $b$ varies, but $\nu$ is the same. The
last plot (d) corresponds to the limit case when
(\ref{realperiodic2}) is reduced to (\ref{nsn}) with $m=1$. }
\label{FigBro}
\end{figure}

The six band-edge states can be
written easily in terms of the Jacobi elliptic functions
\begin{eqnarray}
    &\psi_{0}=\begin{pmatrix}
    { i {\rm dn}\,(x+b/2) \cr  -{\rm dn}\, (x-b/2)}
    \end{pmatrix}\quad, \qquad \psi_{1}=\begin{pmatrix}
    { i {\rm cn}\,(x+b/2) \cr  -{\rm cn}\, (x-b/2)}
    \end{pmatrix}\quad, \qquad \psi_{2}=\begin{pmatrix}
    { i {\rm sn}\,(x+b/2) \cr  -{\rm sn}\, (x-b/2)}
    \end{pmatrix} ,&\nonumber\\
    &\psi_3=\sigma_3\psi_2\quad, \qquad \psi_4=\sigma_3\psi_1\quad,
    \qquad \psi_5=\sigma_3\psi_0\quad,&  \\
   & H_\R \psi_i=E_i \psi_i \quad , \qquad E_{0}=-E_5=-\frac{1}{{\rm sn}\, b} \quad,
    \qquad E_{1}=-E_4=-\frac{{\rm dn}\, b}{{\rm sn}\, b}\quad,
    \qquad E_{2}=-E_3=-\frac{{\rm cn}\, b}{{\rm sn}\, b}\quad.& \nonumber
\end{eqnarray}
From the real nS$_+$ hierarchy we find a non-trivial integral of
motion  $S_3$  which annihilates all the band edge states,
\begin{equation} \label{q3periodic}
    S_3=i\begin{pmatrix}
    {S_3^+ & 0 \cr \  0 & S_3^-}
    \end{pmatrix}\quad, \qquad  [H_\R,S_3]=0\quad,
\end{equation}
where
\begin{eqnarray}
    S_3^+(x)&=&\frac{d^3}{dx^3}+(1+\nu-3\nu{\rm sn}^2\,
    (x+b/2))\frac{d}{dx}+3i\nu{\rm dn}\, (x+b/2){\rm cn}\, (x+b/2){\rm
    sn}\, (x+b/2)\quad, \nonumber\\
    S_3^-(x)&=&S_3^+(x-b)\quad.
\end{eqnarray}
The integral of motion $S_3$  anticommutes with ${\cal R}\sigma_2$,
and we can identify the ${\cal N}=2$ nonlinear supersymmetry. It is
generated by the supercharges
\begin{eqnarray}\label{q3supercharges}
    {\cal Q}_1=S_3\quad,
    \qquad {\cal Q}_2=i{\cal R}\sigma_2 S_3\quad,
\end{eqnarray}
\begin{eqnarray}\label{q3superalgebra}
 [H_\R, {\cal Q}_a]=0 \quad, \qquad \{ {\cal Q}_a, {\cal Q}_b \}
    =2\delta_{ab}P_6(H_\R)\quad, \qquad a,b=1,2\quad,
\end{eqnarray}
where  $P_6(H_\R)$  is a spectral polynomial
\begin{equation}
    P_6(H_\R)=\prod_{k=0}^5 (H_\R-E_k)=
    (H_\R^2-E_0^2)(H_\R^2-E_1^2)(H_\R^2-E_2^2)\quad.
\end{equation}
Another operator from the hierarchy which commutes with $S_3$ but
does not commute with $H_\R$  is
\begin{equation}\label{q2evenper}
    S_2=i\begin{pmatrix}
    {0 &  S_2^- \cr \   S_2^+ & 0}
    \end{pmatrix}\quad, \qquad [S_2,S_3]=0\quad,
\end{equation}
where
\begin{equation}
    S_2^{\pm}=\frac{d^2}{dx^2} \mp\Delta \frac{d}{dx}-
    \frac{1}{2}\left(\Delta^2\mp\ \Delta'-1-\nu+3{\rm ns}^2\,b\right)\quad.
\end{equation}
We can check that $S_2$  satisfies
\begin{eqnarray}
    [H_\R,S_2]=2i\sigma_3\frac{{\rm dn}\,b \,{\rm cn}\,b}{{\rm
    sn}^3\,b}\quad.
    \label{eps1}
\end{eqnarray}

Thus, this example shows how the nonlinear
supersymmetry structure arises for the finite gap BdG
system given by the periodic function (\ref{realperiodic2}). We note here that the Schr\"odinger system (\ref{hisosper}) was also discussed in \cite{samsonov} in the context of Darboux transformations.

\subsubsection{ Real non-periodic solution of the second nS$_+$ equation}

Now we consider the non-periodic limit of (\ref{realperiodic2}) when
the elliptic  parameter $\nu\rightarrow 1$,
\begin{eqnarray}\label{4bandsnon}
    \Delta(x)&=&\coth{b}+\tanh (x-b/2) - \tanh(x+b/2)\quad.
\end{eqnarray}
The main properties of the supersymmetric structure remain invariant
in this limit.  The gap function (\ref{4bandsnon}) is also a
solution of the second nS$_+$ equation
\begin{equation}
    \Delta^{\prime \prime \prime}-6\Delta^{2}\Delta^{\prime}
    +2(1+\frac{3}{\sinh^2 b})\Delta^{\prime}=0\quad.
\end{equation}
The Hamiltonian (\ref{realH}) takes here a form
\begin{equation}\label{h4bandsnon}
    H_\R=i\begin{pmatrix}
    {0 & \frac{d}{dx}+\coth{b}+\tanh (x-b/2) - \tanh(x+b/2)\cr \
    \frac{d}{dx}-(\coth{b}+\tanh (x-b/2) - \tanh(x+b/2) )& 0}
    \end{pmatrix}\quad .
\end{equation}
The square of $H_\R$  produces two self-isospectral reflectionless
Schr\"odinger Hamiltonians,
\begin{equation}
    {\cal H}=H_\R^2=\begin{pmatrix}
    {-\frac{d^2}{dx^2} -\frac{2}{\cosh^2 (x+b/2)} +\coth^2 b & 0
    \cr \  0 & -\frac{d^2}{dx^2} -\frac{2}{\cosh^2 (x-b/2)} +\coth^2 b}
    \end{pmatrix}\quad.
    \label{schisos}
\end{equation}

The spectrum of (\ref{h4bandsnon}) is symmetric with respect to the
zero level of energy. It is composed by two bound states which come
from the shrink of the bands $[E_1,E_2]\rightarrow
\mathcal{E}_1=-\frac{1 }{\sinh b}$, and $[E_3,E_4]\rightarrow
\mathcal{E}_2=\frac{1 }{\sinh b}$,
\begin{equation}
    \psi_{1}=\begin{pmatrix}
    { i\,{\rm sech}\,(x+b/2) \cr -{\rm sech}\, (x+b/2) }
    \end{pmatrix}\quad, \qquad \psi_2=\sigma_3\psi_1\quad,
    \qquad H_\R\psi_{i}= {\mathcal E}_{i}
    \psi_{i} \quad, \qquad i=1,2\quad,
\end{equation}
and by two continuous bands $(-\infty, \mathcal{E}_0]$ and $[
\mathcal{E}_4,\infty)$. The energies in the continuous bands are
doubly degenerate, except the singlet edge values, and  are
described by the eigenfunctions
\begin{equation}
    \psi_{\uparrow}^{\pm}=\begin{pmatrix}
    {\frac{\mp k+i\coth b}{\mathcal{E}_{\uparrow;k}}\mathcal{D}^+_{-1}e^{\pm ik x }
    \cr \mathcal{D}_{-1}^-e^{\pm ik x }}
    \end{pmatrix}\quad, \qquad H_\R\psi_{\uparrow}^{\pm}=
    \mathcal{E}_{\uparrow;k}\psi_{\uparrow}^{\pm} \quad,
    \qquad \mathcal{E}_{\uparrow;k}=\frac{\sqrt{(1+k^2)\sinh^2 b+1}}{\sinh
    b}\quad,
\end{equation}
\begin{equation}
    \psi_{\downarrow}^{\pm}=\sigma_3\psi_{\uparrow}^{\pm}\quad, \qquad
    H_\R\psi_{\downarrow}^{\pm}=\mathcal{E}_{\downarrow;k}\psi_{\downarrow}^{\pm} \quad,
    \qquad
    \mathcal{E}_{\downarrow;k}=-\mathcal{E}_{\uparrow;k}\quad,
\end{equation}
where we introduce the notation
\begin{equation}
    \mathcal{D}_\lambda^{\pm}=\frac{d}{dx}+\lambda \tanh (x\pm b/2)\quad.
\end{equation}
The singlet edge energies $\mathcal{E}_{\uparrow;0}=\mathcal{E}_{4}$
and $\mathcal{E}_{\downarrow;0}=\mathcal{E}_{0}$ correspond to zero
value of the non-negative continuous parameter $k$.

The integral of motion $S_3$ of the corresponding Lax pair
$(H_\R,S_3)$, which annihilates all the singlet states, can be
obtained by taking limit $\nu \rightarrow 1$ in (\ref{q3periodic}).
It has the following structure:
\begin{equation}
    S_3= i\begin{pmatrix} {
    \mathcal{D}^+_{-1}\mathcal{D}_{0}^+\mathcal{D}_{1}^+ & 0  \cr 0 &
    \mathcal{D}^-_{-1}\mathcal{D}_{0}^-\mathcal{D}_{1}^- }
    \end{pmatrix}\quad.
\end{equation}
As in the periodic case we have
\begin{equation}
    [H_\R, \Gamma]=0 \quad, \qquad \{S_3,
    \Gamma\}=0\quad, \qquad \Gamma=\sigma_2{\cal R}\quad,
\end{equation}
and the supersymmetry has the same non linear structure
(\ref{q3supercharges}), (\ref{q3superalgebra}) with the spectral
polynomial changed for
\begin{equation}
    P_6(H_\R)=(H_\R^2-{\mathcal E}_{0}^2)
    (H_\R^2-{\mathcal E}_1^2)^2\quad.
\end{equation}
The components of the operator $S_2$ from (\ref{q2evenper}) reduce
in the infinite period limit to
\begin{eqnarray}
  S_2^{\pm}&=&\frac{d^2}{dx^2}\mp(\coth{b}+\tanh (x-b/2)
    -\tanh(x+b/2))\frac{d}{dx}+\frac{1}{\sinh^2 b}+{\rm sech}^2\, (x
    \pm b/2)\quad .
\end{eqnarray}
This operator obeys the relations
\begin{equation}
    [S_2, H_\R]=2i\frac{\cosh b}{ \sinh^3 b}\sigma_3\quad,
    \label{eps2}
\end{equation}
\begin{equation}
     \quad [S_2,S_3]=0\quad.
     \label{s2s3}
\end{equation}

\section{Complex case}

Here we describe a one-parametric generalization of the finite-gap
systems from Section \ref{IIIA}, given by a complex $\Delta$.  In
such a case an energy  reflection symmetry is partially broken,
but the supersymmetric structure of the BdG system is preserved in
a form to be coherent with the partial discrete symmetry breaking.
A kind of duality underlying the supersymmetric structure is
observed.

\subsection{Complex non-periodic solution of the $(2m-1)^{\rm th}$ nS$_{+}$ equation}

First, consider a simpler non-periodic limit case. We take the
following complex generalization of the kink (\ref{npt}), the
twisted kink \cite{shei,bd2},
\begin{equation}
    \Delta(x;\theta)= m \, \frac{\cosh\left(\,\sin\left(\theta/2\right)\,
     x-i\theta/2\right)}{\cosh\left(\,\sin\left(\theta/2 \right)\,
      x\right)}\, e^{i\theta/2}\quad ,
\label{complex-kink}
\end{equation}
where $m\in \mathbb{N}$, and $\theta$ is a real parameter. The
function $\Delta(x)$ in (\ref{complex-kink}) satisfies the complex
$(2m-1)^{\rm th}$ nS$_{+}$ equation. It obeys relations
\begin{equation}\label{Del+-}
    \Delta(x;\theta+2\pi)=\Delta(-x;\theta)\quad ,\qquad
    \Delta(x,\theta+4\pi)=\Delta(x;\theta)\quad,
\end{equation}
and reduces to the real kink case (\ref{npt}) at $\theta=\pi$. It
takes constant values $+m$ and $-m$ at $\theta=0$ and
$\theta=2\pi$ respectively, when the BdG  system reduces to a free
(1+1)D Dirac particle of mass $m$.

In correspondence with (\ref{Del+-}), the BdG Hamiltonian
(\ref{ham}) anticommutes with the following two mutually commuting
operators
\begin{eqnarray}\label{G12}
    &{\cal G}_1=- U(\frac{\theta}{2})
    \gamma^0T_{2\pi}U^{-1}(\frac{\theta}{2})\quad,\qquad
    {\cal G}_2= U(\frac{\theta}{2}){\cal R}
    T_{2\pi}U^{-1}(\frac{\theta}{2})
    \quad.&
\end{eqnarray}
That is, $\{H,{\cal G}_i\}=0$, $i=1,2$, where $U$ is the operator
of the chiral rotation defined in (\ref{chiralrot}),
 and $T_{2\pi}$ is an operator of the half-period translation
of $\theta$, $T_{2\pi}:\theta\rightarrow \theta+2\pi$. Notice that
the treatment of the operators (\ref{G12}) in complex case
$\theta\neq \pi$ implies the inclusion into consideration of a
``dual" BdG system with shifted  in $2\pi$ value of the parameter
$\theta$. Because of the presence of the operator $T_{2\pi}$ in
(\ref{G12}), the reflection energy symmetry of the real case, as we
shall see below, is partially broken here. On the other hand, the
half-period translation $T_{2\pi}$ disappears in the composition of
operators (\ref{G12}),
\begin{eqnarray}\label{G1R}
    &\Gamma={\cal R}
    U(\frac{\theta}{2})\gamma^5U^{-1}(\frac{\theta}{2})\quad,&
\end{eqnarray}
which is a nonlocal integral of motion of the system,
$[\Gamma,H]=0$.

 We present the BdG Hamitonian (\ref{ham}) in a more
suitable form applying a unitary  transformation given by the matrix
\begin{equation}
    \mathcal{U}_{\theta}=\frac{i}{\sqrt{2}}\begin{pmatrix}
    {e^{-i\frac{\theta}{4}} &-e^{i\frac{\theta}{4}}  \cr \
    -e^{-i\frac{\theta}{4}} & -e^{i\frac{\theta}{4}}}
    \end{pmatrix}\quad,
\label{unitary2}
\end{equation}
which reduces  to (\ref{unitary1})  when $\theta=\pi$. Equivalently,
this corresponds to a choice of the following $\theta$-dependent
representation for $\gamma$-matrices,
\begin{eqnarray}
    \gamma^0=\begin{pmatrix}
    {-i\sin \frac{\theta}{2} & \cos \frac{\theta}{2}   \cr \ -
    \cos \frac{\theta}{2} & i\sin \frac{\theta}{2} }
    \end{pmatrix}
    \quad , \quad
    \gamma^1=\begin{pmatrix}
    {-\cos \frac{\theta}{2} & i\sin \frac{\theta}{2}
    \cr \ -i\sin \frac{\theta}{2} & \cos \frac{\theta}{2} }
    \end{pmatrix}
    \quad , \quad
    \gamma^5=-\sigma_1 \quad.
    \label{gamma-2}
\end{eqnarray}
In this representation the BdG Hamiltonian (\ref{ham}) acquires
the form
\begin{equation}
    H_\C=i\begin{pmatrix}
    { i m \cos \frac{\theta}{2}   & \frac{d}{dx} + m \sin
    \frac{\theta}{2} \tanh (\sin \frac{\theta}{2} x) \cr \
     \frac{d}{dx} - m \sin \frac{\theta}{2} \tanh (\sin
     \frac{\theta}{2} x) & - i m \cos \frac{\theta}{2}  }
    \end{pmatrix}=\begin{pmatrix}
    { - m \cos \frac{\theta}{2}   & i\hat{\mathcal{D}}_{m}
    \cr \  i\hat{\mathcal{D}}_{-m}  &  m \cos \frac{\theta}{2}
    }\quad.
\label{HCnonperiodic}
\end{pmatrix}
\end{equation}
In (\ref{HCnonperiodic})  we introduced the same type of operators
as in the previous section,
\begin{equation}\label{dhat}
    \hat{\mathcal{D}}_{\lambda}=\frac{d}{dx}+\lambda \sin
    \frac{\theta}{2}\,  \tanh \left(\sin \frac{\theta}{2}
    x\right)\quad,
\end{equation}
which satisfy the relation
\begin{eqnarray}\label{Ddilat}
    &\hat{\mathcal{D}}_{\lambda}(x)=\sin\frac{\theta}{2}
    \, \mathcal{D}_{\lambda}(\sin\frac{\theta}{2}x)&
\end{eqnarray}
and reduce to $\mathcal{D}_{\lambda}$, defined in (\ref{unipt}),
 when $\theta=\pi$.
The operators defined in (\ref{G12}) take in representation
(\ref{gamma-2}) the form
\begin{equation}
    {\cal G}_1=\sigma_3T_{2\pi}\quad,
    \qquad {\cal G}_2= {\cal R}T_{2\pi}\quad,
    \label{G12rep}
\end{equation}
and  the nonlocal integral (\ref{G1R}) reduces to (\ref{GammaR}),
$\Gamma={\cal R }\sigma_3$.

Hamiltonian (\ref{HCnonperiodic}) satisfies the relation
$T_{2\pi}H_\C=(H_\C+2m\cos\frac{\theta}{2}\sigma_3)T_{2\pi}$, that
transforms into $[T_{2\pi},H_\C]=0$ in the case $\theta=\pi$. Its
eigenvalues depend on $\theta$ and satisfy the relation ${\cal
E}(\theta+4\pi)={\cal E}(\theta)$ in correspondence with
(\ref{Del+-}). Since $H_\C$ anticommutes with operators
(\ref{G12rep}), we find particularly that if a state $\psi_{\cal
E(\theta)}(x)$ is an eigenstate of (\ref{HCnonperiodic}) of
eigenvalue ${\cal E}(\theta)$, then a state ${\cal
R}T_{2\pi}\psi_{\cal E(\theta)}(x)$ is an eigenstate of eigenvalue
$-{\cal E}(\theta+2\pi)$,
\begin{equation}\label{Etheta2pi}
    H_\C\psi_{\cal
    E(\theta)}(x)={\cal E}(\theta)\psi_{\cal
    E(\theta)}(x)\quad\Rightarrow\quad
    H_\C{\cal R}T_{2\pi}\psi_{\cal E(\theta)}(x)=-{\cal E}(\theta+2\pi)
    {\cal R}T_{2\pi}\psi_{\cal E(\theta)}(x)\quad.
\end{equation}

Explicit form of the eigenstates and eigenvalues of the BdG
Hamiltonian (\ref{HCnonperiodic}), discussed in  Appendix, shows
that for $\theta\neq 0 \, (mod \, \pi)$, if  ${\cal E}(\theta)$ is
an energy of a physical state $\psi_{{\cal E}(\theta)}(x)$ to be
different from the middle bound state, then the state
$\sigma_3T_{2\pi}\psi_{\cal E(\theta)}(x)$ (or, ${\cal
R}T_{2\pi}\psi_{\cal E(\theta)}(x)$) possesses the energy $-{\cal
E}(\theta)$. Together with (\ref{Etheta2pi}), this means that all
the energy levels except the energy of the middle bound state are
periodic functions of $\theta$ with the period $2\pi$.

In more detail, as in the real case, for $\theta\neq 0,2\pi$ the
spectrum of (\ref{HCnonperiodic}) contains $(2m+1)$ singlet states,
$(2m-1)$ of which correspond to bound states. The middle bound state
and its energy are
\begin{equation}\label{middle}
    \psi_{m,0}(x;\theta)=\begin{pmatrix}
    {0 \cr {\rm sech}^m\,(\sin \frac{\theta}{2} x)
    }\end{pmatrix}\quad,
    \qquad \mathcal{E}_m(\theta)=m
    \cos \frac{\theta}{2}\quad,
    \qquad H_\C \psi_{m,0}(x;\theta)= \mathcal{E}_m(\theta)
    \psi_{m,0}(x;\theta)\quad.
\end{equation}
Note that the energy $\mathcal{E}_m (\theta)=m\cos
\frac{\theta}{2}\neq 0$ for $\theta\neq \pi$, and that it varies
monotonically from $+m$ to $-m$ when  $\theta$ varies in the
interval $(0,2\pi)$. For boundary values $\theta=0,2\pi$ the bound
state $\psi_{m,0}(x;\theta)$ as well as all other bound states
disappear, and the system has a spectrum of a free massive (1+1)D
Dirac particle consisting of two symmetric scattering bands
$(-\infty,-m]$ and $[m,\infty)$. Thefore,  for $\theta\neq 0\,
(mod\, \pi)$ the energy reflection symmetry of the real case is
partially broken.

Relations (\ref{Etheta2pi}) and (\ref{middle}) show that the
middle bound states of the two BdG systems with parameter $\theta$
shifted in $2\pi$ satisfy the relation ${\cal E}_m(\theta)=-{\cal
E}_m(\theta+2\pi)$. This equality together with (\ref{Etheta2pi})
correspond to a kind of duality relation
\begin{equation}
    H_\C(x;\theta)=-H_\C(-x;\theta+2\pi)\quad,
    \label{H+2pi}
\end{equation}
which is reflected in the structure of the spectrum shown on Fig.
\ref{FigComN}, cf. \cite{duality}.

\begin{figure}[h!]
\centering
\label{FigComN1}
\subfigure[$m=2$] {
\includegraphics[scale=0.8]{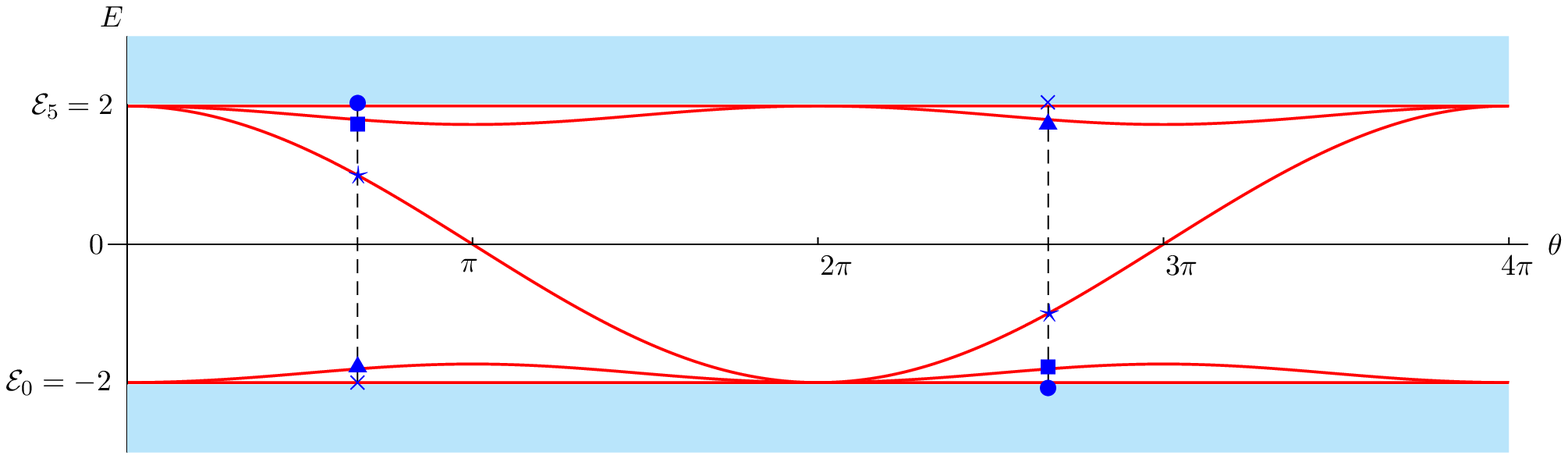}}
\label{FigComN2} 
\subfigure[$m=3$]
{\includegraphics[scale=0.8]{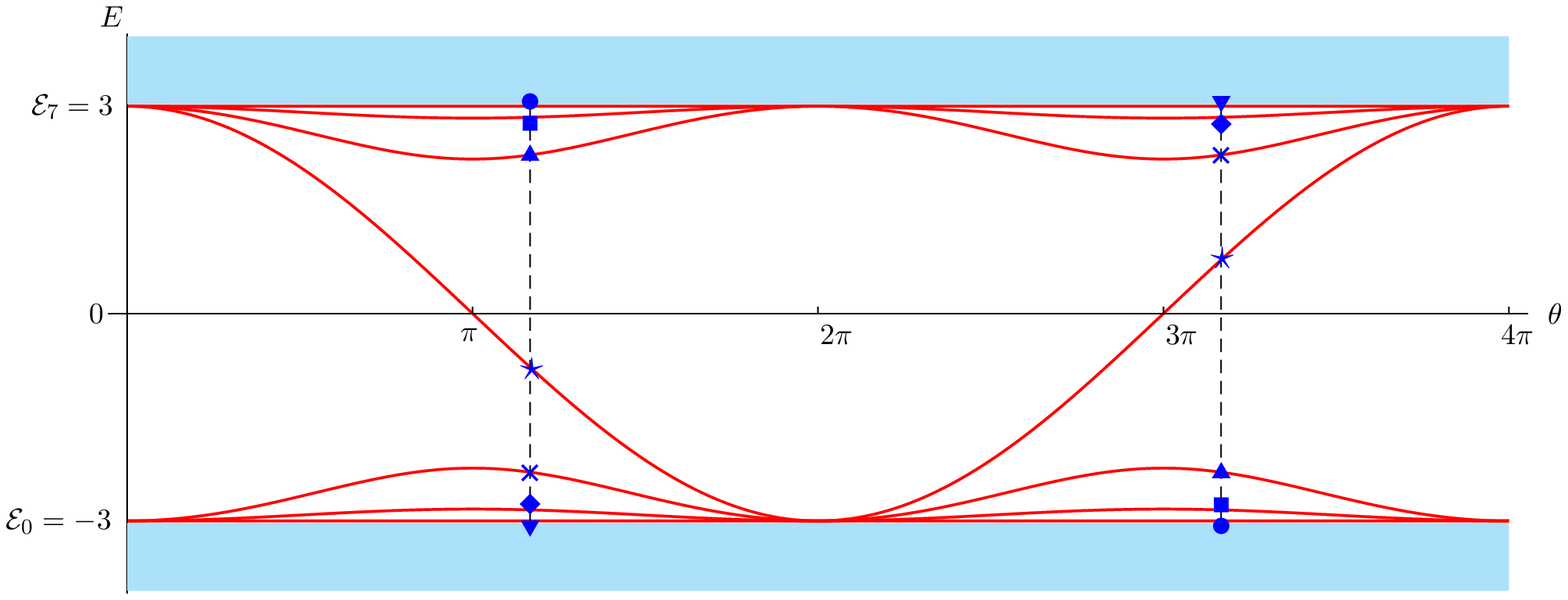}}
\caption{Spectrum of the BdG system (\ref{HCnonperiodic}) for $m=2$
and $m=3$. The red lines correspond to the singlet states energies;
the middle red line reveals a partial breaking of the energy
reflection symmetry. The pairs of the same symbols on the vertical
lines indicate singlet energies for dual systems with shifted in
$2\pi$ parameter $\theta$.} \label{FigComN}
\end{figure}

Observe that
\begin{equation}
    H_\C^2=\begin{pmatrix}
    { -\frac{d^2}{dx^2} -\frac{m(m-1)\sin^2 \frac{\theta}{2} }{\cosh^2 (\sin
    \frac{\theta}{2} x)} +m^2  & 0 \cr \ 0 &  -\frac{d^2}{dx^2} -\frac{m(m+
   1)\sin^2 \frac{\theta}{2} }{\cosh^2 (\sin
    \frac{\theta}{2} x)} +m^2  }
    \end{pmatrix}
    \label{Hnonperiodic}
\end{equation}
is diagonal, with Schr\"odinger-type Hamiltonians of
P\"oschl-Teller form on the diagonal. In the general complex case
of the BdG Hamiltonian, this \emph{local} diagonalization of $H^2$
is not possible.

The non-trivial integral of motion which comes from the Lax pair
relation of the $n$S$_{+}$ hierarchy can be presented in the form
\begin{eqnarray}
    S_{2m}&=&\begin{pmatrix}
    {2mi (-1)^m \cos \frac{\theta}{2} \hat{\mathcal{D}}_{-m+1}^{2m-1}
    & ( \hat{\mathcal{D}}_{-m}^{2m} )^{\dagger} \cr \ \hat{\mathcal{D}}_{-m}^{2m}  & 0}
    \end{pmatrix} \nonumber\\ &=&\begin{pmatrix}
    {2mi (-1)^m \cos \frac{\theta}{2} \hat{\mathcal{D}}_{-m+1}
    \hat{\mathcal{D}}_{-m+2}...\hat{\mathcal{D}}_{m-1} &
    \hat{\mathcal{D}}_{-m+1}\hat{\mathcal{D}}_{-m+2}...
    \hat{\mathcal{D}}_{m-1}\hat{\mathcal{D}}_{m}
    \cr \
    \hat{\mathcal{D}}_{-m}\hat{\mathcal{D}}_{-m+1}...
    \hat{\mathcal{D}}_{m-2}\hat{\mathcal{D}}_{m-1}
    & 0}
\end{pmatrix}\quad.\label{S2mc}
\end{eqnarray}
Since nonlocal operator $\mathcal{R}\sigma_3$ commutes with the
Hamiltonian (\ref{HCnonperiodic}) and anticommutes  with
$S_{2m}$},
\begin{eqnarray}
    [\mathcal{R}\sigma_3,H_\C]=0\quad, \qquad
    \{\mathcal{R}\sigma_3,S_{2m}\}=0\quad,
\end{eqnarray}
we identify the operators
\begin{equation}\label{Qcomplex}
    \mathcal{Q}_1=S_{2m}\quad,
    \qquad \mathcal{Q}_2=i\mathcal{R}\sigma_3\mathcal{Q}_1
\end{equation}
as the supercharges. They generate the ${\cal N}=2$ polynomial
superalgebra of the form (\ref{superalgebra}),
\begin{eqnarray}
    [\mathcal{Q}_a,H_\C]=0\quad,
    \qquad \{\mathcal{Q}_a,\mathcal{Q}_b\}=
    2\delta_{ab}P_{4m}(H_\C)\quad,&
    \label{superalgebraC}
\end{eqnarray}
in which the polynomial (\ref{q2nonperiodic}) becomes
\begin{equation}
    P_{4m}(H_\C)=(H_\C-\mathcal{E}_{0})
    (H_\C-\mathcal{E}_{2m})\prod_{i=1}^{2m-1}
    (H_\C-\mathcal{E}_{i})^2\quad.
\label{q2complexnonperiodic}
\end{equation}

Let us discuss now the nature of supersymmetry of  the associated
extended Schr\"odinger system described by the Hamiltonian
(\ref{Hnonperiodic}). Though (\ref{Hnonperiodic}) is a square of the
BdG Hamiltonian  (\ref{HCnonperiodic}), the latter can not be
treated as a usual supercharge of linear ${\cal N}=2$ supersymmetry
with $\sigma_3$ identified as a $\Z_2$-grading operator. The reason
is that though the quadratic diagonal operator (\ref{HCnonperiodic})
commutes with $\sigma_3$, the linear operator (\ref{HCnonperiodic})
does not anticommute with the diagonal Pauli matrix. In the rest of
this section we present three distinct but  related possible
interpretations of the supersymmetry associated with the second
order matrix diagonal Hamiltonian
\begin{equation}\label{HSHC2}
    {\cal H}_\C=H_\C^2\quad.
\end{equation}

First, we note that the BdG Hamiltonian (\ref{HCnonperiodic}) can be
presented as
\begin{equation}\label{hcthc}
    H_{\C}=H_\R^\theta-m \cos \frac{\theta}{2} \sigma_3\quad,
\end{equation}
where
\begin{equation}
    H_\R^\theta=\begin{pmatrix}
    { 0   & i\hat{\mathcal{D}}_{m}
    \cr \  i\hat{\mathcal{D}}_{-m}  &  0
    }\quad,
\label{Xnota}
\end{pmatrix}
\end{equation}
and $\hat{\mathcal{D}}_{m}$ is defined in (\ref{dhat}). Operator
(\ref{Xnota}) is the integral of motion for the system ${\cal
H}_\C$. In correspondence with (\ref{Ddilat}), it is the rescaled
real BdG Hamiltonian $H_\R(x)$ defined in (\ref{unipt}),
$H^\theta_\R(x)=\sin\frac{\theta}{2}H_\R(\sin\frac{\theta}{2}x)$.
Unlike the BdG Hamiltonian (\ref{hcthc}), it anticommutes with
$\sigma_3$ and so, can be identified as one of two supercharges of
the ${\cal N}=2$ linear supersymmetry, $\hat{\cal Q}_1=H_\R^\theta$.
Defining the second supercharge in a usual way, $\hat{\cal
Q}_2=i\sigma_3\hat{\cal Q}_1$, we  arrive at the superalgebra
\begin{eqnarray}
    &\{\hat{\cal Q}_a,\hat{\cal Q}_b\}=2\delta_{ab}\left(
    {\cal H}_\C-m^2\cos^2\frac{\theta}{2}\right)\quad,\qquad
    [\hat{\cal Q}_a, {\cal H}_\C]=0\quad.&\label{N2central}
\end{eqnarray}
{}Since supercharges $\hat{\cal Q}_a$ are hermitian, from the
first of these relations  and relation (\ref{HSHC2}) we find that
the energies of the finite-gap BdG system (\ref{HCnonperiodic})
satisfy the inequality ${\cal E}^2\geq m^2\cos^2\frac{\theta}{2}$.
Since for $\theta\neq 0\, (mod\, 2\pi)$  system
(\ref{HCnonperiodic}) has an odd number of $2m+1$ singlet states
in its spectrum, the supersymmetric structure (\ref{N2central})
of the associated Schr\"odinger system implies that the reflection
energy symmetry is broken for $\theta\neq \pi\, (mod\, 2\pi)$.

As in the real case $\theta=\pi$, relation (\ref{HSHC2}) means that
the BdG nontrivial integral (\ref{S2mc}) is also the integral of
motion for the  Schr\"odinger system ${\cal H}_\C$. Like the BdG
Hamiltonian $H_\C$, integral $S_{2m}$ does not have a definite
$\Z_2$-parity with respect to the grading operator $\sigma_3$, but
analogously to (\ref{hcthc}) it also can be decomposed into bosonic
(diagonal) and fermionic (antidiagonal) operators
\begin{equation}\label{S2mYth}
    S_{2m}={\cal F}^\theta+ {\cal B}^\theta\quad.
\end{equation}
${\cal F}^\theta$ is a rescaled nontrivial integral  of a real
case, ${\cal
F}^\theta(x)=\left(\sin\frac{\theta}{2}\right)^{2m}S_\R(\sin\frac{\theta}{2}x)$,
with $S_\R(x)$ denoting here the operator (\ref{Qnon}). Bosonic
operator is
\begin{equation}\label{Bcal}
    {\cal B}^\theta=2m(-1)^m\cos\frac{\theta}{2}i\hat{\cal
    D}^{2m-1}_{-m+1}\Pi_+\quad ,
\end{equation}
where $\Pi_+=\frac{1}{2}(1+\sigma_3)$ is a projector. Both operators
${\cal F}^\theta$ and ${\cal B}^\theta$ are the integrals of motion
for the Schr\"odinger system (\ref{HSHC2}). Bosonic integral ${\cal
B}^\theta$ can be related to a rescaled operator (\ref{BdGtri}) with
index $m$ changed for $m-1$. One can find a superlagebra generated
by the  fermionic set of integrals $\hat{\cal Q}_a$, ${\cal
S}_1={\cal F}^\theta$, ${\cal S}_2=i\sigma_3{\cal S}_1$, and by the
bosonic integral ${\cal B}_\theta$ together with the Hamiltonian
(\ref{HSHC2}). It is a nonlinear extension of superalgebra
(\ref{N2central}), whose explicit form we do not display here.

Instead of hermitian integrals $\hat{\cal Q}_a$, $a=1,2$, associated
with decomposition (\ref{hcthc}), consider now a pair of hermitian
conjugate integrals
\begin{equation}
Q=\begin{pmatrix}
    {-\frac{m}{\sqrt{2}}\cos \frac{\theta}{2} & 0 \cr
    \ \hat{D}_{-m} & \frac{m}{\sqrt{2}}\cos \frac{\theta}{2}}
    \end{pmatrix}\quad, \qquad Q^{\dagger}=\begin{pmatrix}
    {-\frac{m}{\sqrt{2}}\cos \frac{\theta}{2} & -\hat{D}_{m}
    \cr \ 0 & \frac{m}{\sqrt{2}}\cos \frac{\theta}{2}}\quad,
    \end{pmatrix}
    \label{qq+}
\end{equation}
which are certain  linear combinations of the integrals $\hat{\cal
Q}_1$, $\hat{\cal Q}_2=i\sigma_3\hat{\cal Q}_1$ and $\sigma_3$.
Together with the Schr\"odinger Hamiltonian ${\cal H}_\C$ they
generate the centrally extended ${\cal N}=2$ superalgebra
\begin{eqnarray}\label{s1}
    \{Q,Q^{\dagger}\}={\cal H}_\C\quad,\qquad
    [{\cal H}_\C,Q] =[{\cal H}_\C,Q^{\dagger}]=0\quad,
 \end{eqnarray}
 \begin{eqnarray}\label{s2}
     \{Q,Q\}= \{Q^{\dagger},Q^{\dagger}\}={\cal C}\quad,
\end{eqnarray}
in which the operator
\begin{equation}\label{Ccal}
     {\cal C}=m^2 \cos^2
     \frac{\theta}{2}I_2\quad,
\end{equation} proportional to the $2\times 2$ unit matrix $I_2$,
plays the role of the central charge. Relation ${\cal H}_\C\geq
{\cal C}$ is implied by superalgebra (\ref{s1}), (\ref{s2}) via
relations  $(H_R^\theta)^2\geq 0$, $H_\R^\theta=i(Q-Q^\dagger)$.

Central extension of ${\cal N}=2$  superalgebra of such a structure
was discussed in \cite{Spector} within the framework of
supersymmetric quantum mechanics. Since  mutually conjugate matrix
integrals (\ref{qq+}) do not anticommute with $\sigma_3$, they can
not be identified as supercharges if $\sigma_3$ is treated as the
$\Z_2$-grading operator $\Gamma$. In our case, however, we can
identify one of the operators (\ref{G12rep}) as $\Gamma$. They both
commute with ${\cal H}_\C$, and identify the Schr\"odinger
Hamiltonian ${\cal H}_\C$ and the integral $\sigma_3$ as the bosonic
operators, while integrals (\ref{qq+}) are identified by any of them
as the fermionic operators. Having in mind that
$[\sigma_3,Q]=-2iQ^\dagger$, $[\sigma_3,Q^\dagger]=2iQ$, the
operator $\sigma_3$ can be treated then as a generator of
$R$-symmetry.

Finally, let us discuss the  tri-supersymmetric structure of the
system (\ref{HSHC2}). From the described properties of the
finite-gap BdG system (\ref{HCnonperiodic}) it follows that the
involutive operators (\ref{G12rep}) and their composition ${\cal
R}\sigma_3$ are the trivial integrals of motion for the associated
Schr\"odinger system (\ref{HSHC2}), $[{\cal G}_i,{\cal
H}_\C]=[{\cal R}\sigma_3,{\cal H}_\C]=0$. Let us denote ${\cal
X}^\theta=H_\C$ and ${\cal Y}^\theta=S_{2m}$, and introduce the
operator
\begin{equation}\label{zth}
    {\cal Z}^\theta=\frac{1}{2}
    \{{\cal X}^\theta,{\cal Y}^\theta\}\quad.
\end{equation}
Operators ${\cal X}^\theta$, ${\cal Y}^\theta$ and ${\cal
Z}^\theta$ are nontrivial integrals for the Schr\"odinger system
(\ref{HSHC2}), which at $\theta=\pi$ reduce to the integrals of
the Schr\"odinger system corresponding to the real case. The
choice of one of the three trivial integrals as the $\Z_2$-grading
operator classifies the nontrivial integrals as bosonic and
fermionic operators in correspondence with relations
\begin{equation}\label{Xgra}
    [ {\cal X}^\theta,{\cal R}\sigma_3]=0\quad,
    \qquad \{{\cal
    Y}^\theta,{\cal R}\sigma_3\}=\{{\cal Z}^\theta,{\cal
    R}\sigma_3\}=0\quad,
\end{equation}
\begin{equation}\label{Ygra}
    [ {\cal Y}^\theta,{\cal R}T_{2\pi}]=0\quad, \qquad \{{\cal X}^
    \theta,{\cal R}T_{2\pi}\}=\{{\cal Z}^\theta,{\cal
    R}T_{2\pi}\}=0\quad,
\end{equation}
\begin{equation}\label{Zgra}
    [ {\cal Z}^\theta,\sigma_3 T_{2\pi} ]=0 \quad, \qquad \{{\cal
    X}^\theta,\sigma_3 T_{2\pi}\}=\{{\cal Y}^\theta,
    \sigma_3 T_{2\pi}\}=0\quad.
\end{equation}

As we noted above, for $\theta\neq \pi$ the treatment of the
operators ${\cal R}T_{2\pi}$ and $\sigma_3 T_{2\pi}$ requires a
consideration of the pair of the corresponding dual systems.

\subsection{Complex periodic solution of the $(2m-1)^{\rm th}$ nS$_{+}$ equation}

A periodic generalization of the function $\Delta(x)$ in
(\ref{complex-kink}) that also satisfies the $(2m-1)^{\rm th}$
nS$_{+}$ equation and includes as special case the examples (real
and complex) discussed above is [the $m=1$ case was studied in
\cite{bd2, bdt}]
\begin{eqnarray}
    \Delta(x;\theta)&=&-m A \frac{\sigma(A\, x+i{\bf K}^\prime
    -i\theta/2)}{\sigma(A\, x+i{\bf K}^\prime)\sigma(i\theta/2)}
    \,\exp\left[i A\, x \left(-i\,\zeta(i\theta/2)+i\,{\rm
    ns}(i\theta/2)\right)+i\,\theta \eta_3/2\right] \quad ,
\label{complex-kink-crystal}
\end{eqnarray}
where the parameter $A$ sets the scale, $ A=A( \theta, \nu)\equiv
-2i\, {\rm sc}\left(i\theta/4\right) {\rm
nd}\left(i\theta/4\right)$. The functions
$\sigma=\sigma(x;\omega_1,\omega_2)$ and
$\zeta=\zeta(x;\omega_1,\omega_2)$ are the Weierstrass sigma and
zeta functions \cite{lawden}. The real and imaginary half-periods
are $\omega_1={\bf K}(\nu)$, and $\omega_3=i\,{\bf K}^\prime\equiv i{\bf K}(1-\nu)$.
The constant $\eta_3\equiv\zeta(i{\bf K}^\prime)$ is purely imaginary.
The parameter $\theta$ plays the same role as in the previous
non-periodic case, but now its period interval is $[0,8{\bf K}'(\nu)]$.
Using the properties of elliptic functions one can check that for
the specific values of the parameter $\theta=2{\bf K}'$ and
$\theta=6{\bf K}'$, (\ref{complex-kink-crystal})  reduces to the real
periodic solution (\ref{nsn}). In the infinite period limit $\nu
\rightarrow 1$, the solution (\ref{complex-kink-crystal}) reduces
to the complex non-periodic case (\ref{complex-kink}). Meanwhile
for $\theta =0$, $\theta=4{\bf K}^\prime$ or $\theta=8{\bf
K}^\prime$, the solution reduces to a free particle case given by
$\Delta=b_1 e^{i b_2 x}$, where $b_i$, $i=1,2$ are constants, that
produces a simple shift in the energy \cite{bd2}.

To analyze the properties of function
(\ref{complex-kink-crystal}), we decompose it in terms of the
amplitude and phase; $\Delta=Me^{i \chi}$,
\begin{eqnarray}
    M(x)&=&|\Delta|=m A \sqrt{\mathcal{P}(Ax+i{\bf K}^\prime)-\mathcal{P}(i
    \theta/2)}\quad,\\
    \chi(x)&=&A(-i \zeta(i \theta/2)+i {\rm ns}(i
    \theta/2))x+\frac{i}{2}\ln \left(\frac{\sigma(Ax+i{\bf K}'+i
    \theta/2)}{\sigma(Ax+i{\bf K}'-i
    \theta/2)}\right)+\frac{\eta_3\theta}{2}+\frac{\pi}{2}\quad ,
\end{eqnarray}
where $\mathcal{P}(x)$ is the elliptic Weierstrass function.
Meanwhile $M$ is a function of period $T=2{\bf K}/A$ in $x$, $\chi$
changes in $\varphi=2{\bf K}(-i \zeta(i \theta/2)+i {\rm ns}(i
\theta/2)-\frac{\eta \theta}{2{\bf K}} )$ over the period $T$. As a result
(\ref{complex-kink-crystal}) changes as $\Delta(x+2{\bf K}/A)=e^{i\varphi}
\Delta(x)$, , i.~e. (\ref{complex-kink-crystal}) is a quasi-periodic
in $x$ function. Hence, in correspondence with (\ref{chiralrot}),
the BdG Hamiltonian (\ref{ham-bdg}) satisfies the relation
\begin{equation}
    H(x+T)=e^{i\gamma_5 \varphi/2} H(x) e^{-i\gamma_5 \varphi/2}
    \quad,
    \label{chiral-rotation}
\end{equation}
i. e. it is periodic up to a chiral transformation. This means that
$H(x+T)$ and $H(x)$  systems are unitary equivalent, and the BdG
Hamiltonian with quasi-periodic complex function
(\ref{complex-kink-crystal}) describes the finite-gap periodic
system.

\begin{figure}[h!]
\centering
\includegraphics[scale=0.835]{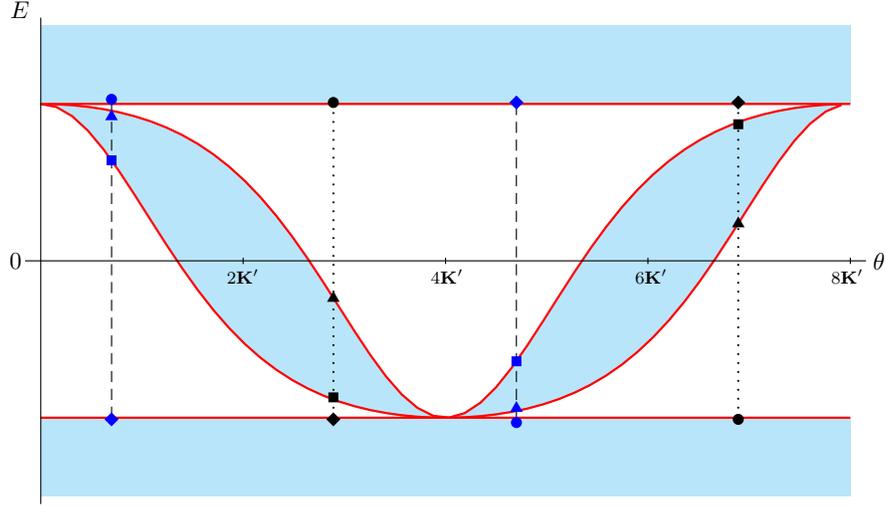}
\caption{Spectrum of the BdG system for the complex periodic
solution (\ref{complex-kink-crystal}) with $m=1$ in dependence on
$\theta$. The pairs of the same symbols on the red lines
corresponding to the singlet edge-states reflect a symmetry
associated with the duality between the pairs of the systems with
parameter $\theta$ shifted in the half-period $4{\bf K}'(\nu)$.}
\label{FigComp}
\end{figure}

The spectrum of this system shown on Figure \ref {FigComp} has the
properties similar to those we described in the non-periodic case.
The continuum bands $(-\infty, E_0]$ and $[E_{4m-1}, \infty)$, as
well as any corresponding pair from $2m-2$ valence bands  which do
not include the middle band are symmetric with respect to the zero
energy level. As in the non-periodic case, the spectrum reveals
also a symmetry associated with a duality relation, which, up to a
unitary chiral transformation, has the form of (\ref{H+2pi}) with
$2\pi$ changed here for the $\theta$-half-period equal to
$4{\bf K}'(\nu)$ \footnote{We note that the spectrum also reveals a symmetry 
$E(2{\bf K}'-\theta)\rightarrow - E(2{\bf K}'+\theta)$ [and
${\cal E}(\pi-\theta)\rightarrow -{\cal E}(\pi+\theta)$ in the
non-periodic case]. This symmetry is not independent, but is a 
consequence of the duality relation and the invariance of the
spectrum under reflection with respect to $\theta=4{\bf K}'$
[$\theta=2\pi$]. The latter, in turn, follows from invariance of
the Hamiltonian (\ref{HCperiodic}) [(\ref{HCnonperiodic})] under
the change $\theta\rightarrow -\theta$ and its quasi-invariance
[invariance] under the shift for the period $8{\bf K}'$ [$4\pi$].}.

To reveal the ${\cal N}=2$ nonlinear supersymmetric structure and
the partial breaking of the energy reflection symmetry in the
finite-gap BdG system given by (\ref{complex-kink-crystal}), we
identify first the $\Z_2$-grading operator. To this aim we note that
the imaginary part of (\ref{complex-kink-crystal}) is an \emph{even}
function, while its real part is  \emph{parity-odd}. Applying the
unitary transformation  (\ref{unitary2}) with parameter $\theta=\pi$
to the BdG Hamiltonian (\ref{ham-bdg}) we obtain
\begin{equation}
    H_\C=\begin{pmatrix} {\frac{i}{2}(\Delta
    -\Delta^* )& i\frac{d}{dx}+
    \frac{i}{2}(\Delta^* +\Delta
    ) \cr \ i\frac{d}{dx}- \frac{i}{2}(\Delta^*
    +\Delta) &
  -\frac{i}{2}(\Delta
    -\Delta^* )}\end{pmatrix}=\begin{pmatrix} {-\text {Im}\,\Delta& i
    \left( \frac{d}{dx}+\text {Re}\,\Delta \right) \cr \ i
    \left( \frac{d}{dx}-\text {Re}\,\Delta \right) &
 \text {Im}\,\Delta }\end{pmatrix} \quad.
\label{HCperiodic}
\end{equation}
The diagonal and anti-diagonal parts of the  transformed BdG
Hamiltonian  have opposite parities, and similarly to the
antiperiodic case, (\ref{HCperiodic}) commutes with the nonlocal
operator $\mathcal{R}\sigma_3$, $[\mathcal{R}\sigma_3,H_\C]=0$. We
can also find the analogs of the the operators (\ref{G12}) which
factorize the integral $\mathcal{R}\sigma_3$ and anticommute  with
(\ref{HCperiodic}). These are
\begin{equation}
    {\cal G}_1=\sigma_3 T_{4{\bf K}'}\quad, \qquad {\cal G}_2={\cal R} T_{4{\bf K}'} \quad ,
\end{equation}
\begin{equation}
    \{{\cal G}_i,H_\C \}=0\quad, \qquad i=1,2 \quad.
\end{equation}
Here we introduced the half-period $\theta$-translation operator
$T_{4{\bf K}'}:\theta \rightarrow \theta + 4{\bf K}'$.

 As in the previous
non-periodic case, from relation (\ref{q}) we can construct an
operator $S_{2m}$ which is a nontrivial integral of motion $[H_\C,
S_{2m}]=0$. In  contrast with the previous cases, here it is not
easy to find a closed general form for this operator, and we just
refer to Eq. (\ref{q}). Conserved quantity $S_{2m}$ anti-commutes
with integral $\mathcal{R}\sigma_3$, that allows us to identify
the integrals $\mathcal{Q}_1=S_{2m}$ and
$\mathcal{Q}_2=i\mathcal{R}\sigma_3\mathcal{Q}_1$ as supercharges.
They generate the nonlinear ${\cal N}=2$ supersymmetry,
\begin{eqnarray}
    [\mathcal{Q}_a,H_\C]=0\quad,
    \qquad \{\mathcal{Q}_a,\mathcal{Q}_b\}=
    2\delta_{ab}P_{4m}(H_\C)\quad,&
    \label{superalgebraC-2}
\end{eqnarray}
where
\begin{equation}
    P_{4m}(H_\C)=\prod_{i=1}^{4m-1}(H_\C-E_{i})\quad,
\label{q2complexperiodic}
\end{equation}
is the spectral polynomial, and  $E_i$ are the band-edge energies.
In the  limit $\nu\rightarrow 1$, (\ref{q2complexperiodic})
reduces to the polynomial  (\ref{q2complexnonperiodic}) of the
non-periodic case.

The square of the BdG Hamiltonian (\ref{HCperiodic}) takes here a
non-diagonal form in contrast with the diagonal form of the
associated Schr\"odinger Hamiltonian we had in the non-periodic
case.  Explicitly, we find that
\begin{equation}\label{hc2}
    {\cal H}_\C=H_\C^2=\begin{pmatrix} { -\frac{d^2}{dx^2}+\text
    {Re}^2\,\Delta+(\text {Re}\,\Delta)'+\text {Im}^2\,\Delta & i(
    \text {Im}\,\Delta)' \cr \ - i( \text {Im}\,\Delta)' &
      -\frac{d^2}{dx^2}+\text {Re}^2\,
  \Delta-(\text {Re}\,\Delta)'+\text {Im}^2\,\Delta }\end{pmatrix}
\end{equation}
includes purely imaginary off-diagonal terms. The second order
Hamiltonian (\ref{hc2}) can be written in a more simple form if we
define the matrix superpotential ${\cal W}$,
\begin{equation}
    {\cal W}=\text {Re}\,\Delta \,\sigma_3-\text {Im}\,\Delta\, \sigma_2
    \quad,
    \qquad {\cal H}_\C=-\frac{d^2}{dx^2}+{\cal W}^2+{\cal W}' \quad.
\end{equation}
Though  the second order matrix Hamiltonian includes the
off-diagonal terms, which can be interpreted as a coupling between
the partner subsystems,  the tri-supersymmetric structure still
presents in this case as well. With the identification of operators
${\cal X}^\theta=H_\C$, ${\cal Y}^\theta=S_{2m}$ and ${\cal
Z}^\theta=\frac{1}{2}\{{\cal X},{\cal Y}\}$, as in the limit case
(\ref{zth}) when $\nu\rightarrow 1$, we can treat them as nontrivial
integrals of the second order system ${\cal H}_\C$. Their
fermionic-bosonic nature and the concrete form of the corresponding
superalgebra  depend on which of the three trivial integrals
\begin{equation}\label{gradingC}
{\cal R} \sigma_3\quad, \qquad {\cal R}T_{4{\bf K}'}\quad, \quad {\rm
or}\qquad \sigma_3 T_{4{\bf K}'}
\end{equation}
is identified as the $\Z_2$-grading operator, cf. Eqs.
(\ref{Xgra})--(\ref{Zgra}).

\section{Conclusions}
 The Bogoliubov/de Gennes Hamiltonian (\ref{ham}) appears in a wide variety of contexts
 in theoretical and mathematical  physics. Here we have shown first that the Ginzburg-Landau
 expansion of the associated grand canonical potential has a recursive structure that is precisely
 that of the AKNS integrable hierarchy. This explains why it is possible to solve the corresponding
 inhomogeneous gap equation by a simple ansatz for the Gorkov resolvent, as was done in \cite{bd2}.
 We further have shown that this integrable AKNS hierarchy structure, and its finite gap solutions,
 provide the natural formalism for characterizing nonlinear quantum mechanical supersymmetry
 and tri-supersymmetry. We have illustrated these ideas with explicit  solutions
 in terms of trigonometric and elliptic functions, for which the spectral and quantum
 mechanical supersymmetry properties can be computed in closed-form. Mathematically, 
 there are open questions
 regarding the explicit form of the multi-dimensional theta function solutions to
 the ${\rm AKNS}_N$ hierarchy equations for $N\geq 2$, which are associated with
 hyperelliptic [rather than elliptic] curves, and Riemann surfaces of genus
 $g\geq 2$ \cite{belokolos,gesztesy,gw,enolskii}.
 This is deeply related to the problem  of the reduction of abelian integrals on
 Riemann surfaces \cite{gesztesy,gw,enolskii,babich}.  An interesting case that is not
 yet well understood in such explicit terms is the complex AKNS solution with
 an even number of gaps.
 Physically,  interesting open problems are: (i) the question of the existence
 of self-consistent  solutions $\Delta(x)$ to the inhomogeneous gap equation for
 the {\it massive} ${\rm NJL}_2$ model, for which approximate solutions are known \cite{karbstein},
 but no analytic solutions are known; (ii) the search for self-consistent crystalline solutions
 in higher dimensional models \cite{bringoltz,nickel,DGKL}. It would also be interesting
  to apply these integrability and quantum mechanical supersymmetry
  structures to the generalized Gross-Neveu and
   NJL models coming from string theories \cite{harvey,parnachev,basu,sachs}.

\section{Appendix }

In this Appendix we present the solution of the eigenvalue problem
for the non-periodic BdG Hamiltonian (\ref{HCnonperiodic}) in terms
of solution of the same problem for the real case $\theta=\pi$
(\ref{unipt}). These two problems can be related in a simple way
because of a very special nature of the BdG Hamiltonian
corresponding to the twisted kink (\ref{complex-kink}):  in general
case, in contrast with   (\ref{HCnonperiodic}), the square of
(\ref{ham-bdg}) can not be transformed into a local diagonal form.

Let us denote  (\ref{eigenRB}) and (\ref{eigenRS}) universally by
$\psi_+$ and $\psi_-$ as solutions with positive and negative energy
respectively. As we noted, for zero energy eigenvalue, Eqs.
(\ref{eigenRB}) and (\ref{eigenRS}) give the same eigenstate
(\ref{psim0}). Due to the energy reflection symmetry of the real
case, eigenfunctions of  positive and negative energy sectors are
related as $\psi_-=\sigma_3 \psi_+$. We have
\begin{eqnarray}\label{realBdGnon}
    &H_\R^\theta(x)\tilde{\psi}_{\pm}(x)=\pm \sin \frac{\theta}{2} \tilde{{\cal E}}_+
    \tilde{\psi}_{\pm}(x)\quad,\qquad
    \tilde{\psi}_\pm(x)=\psi_\pm(\sin\frac{\theta}{2}x)\quad,&
\end{eqnarray}
where
$H_\R^\theta(x)=\sin\frac{\theta}{2}H_\R(\sin\frac{\theta}{2}x)$
corresponds to the notation introduced in (\ref{hcthc}), and by
$\tilde{{\cal E}}_+$ we denoted the eigenvalue of the state
$\psi_+$, $H_\R\psi_+=\tilde{{\cal E}}_+\psi_+$. Using relation
(\ref{hcthc}), we can rewrite (\ref{realBdGnon}) as
\begin{eqnarray}
    &\left( H_\C+m\cos \frac{\theta}{2}\sigma_3\right)\tilde{\psi}_+= \sin
    \frac{\theta}{2}\tilde{{\cal E}}_+ \tilde{\psi}_{+}\quad,\qquad
    \left( H_\C+m\cos \frac{\theta}{2}\sigma_3\right)
    \sigma_3\tilde{\psi}_+=- \sin \frac{\theta}{2}\tilde{{\cal E}}_+
    \sigma_3\tilde{\psi}_{+} \quad.&
\end{eqnarray}
{}From here we find the eigenvalues and eigenstates of the
Hamiltonian $H_\C$,
\begin{equation}\label{HC+-}
    H_\C\psi_\pm(x;\theta)=\pm {\cal E}_+(\theta)\psi_\pm(x;\theta)\quad,
\end{equation}
\begin{equation}\label{Epsitheta}
    {\cal E}_+(\theta)=\sqrt{\sin^2
    \frac{\theta}{2}\tilde{{\cal E}}_+^2+m^2\cos^2
    \frac{\theta}{2}}\quad,\qquad
    \psi_\pm(x;\theta)=(\alpha_\pm(\theta)+\sigma_3)\tilde{\psi}_+(x)\quad,
\end{equation}
where
\begin{equation}\label{alpha+}
    \alpha_+(\theta)=\frac{-m\cos
    \frac{\theta}{2}}{{\cal E}_+(\theta)-\sin\frac{\theta}{2}\tilde{{\cal
    E}}_+}\quad,\qquad
    \alpha_-(\theta)=\alpha_+(\theta+2\pi)\quad.
\end{equation}
Note that the middle bound state energy ${\cal
E}_m(\theta)=m\cos\frac{\theta}{2}$, for which $\tilde{\cal
E}_+={\cal E}_+(\pi)=0$ in (\ref{Epsitheta}), is antiperiodic
function of the parameter $\theta$, ${\cal E}_m(\theta+2\pi)=-{\cal
E}_m(\theta)$. For the rest of the spectrum, energy levels change
periodically in $\theta$ with the period $2\pi$, that is reflected
coherently in Fig. \ref{FigComN}.

\section{Acknowledgments}
GD thanks the DOE for support under grant DE-FG02-92ER40716, and
is particularly grateful to F. Gesztesy and V. Enolskii for their
valuable comments. FC acknowledges the support from  CONICYT,  and
thanks the Physics Department of the University of Connecticut
for support and hospitality during an extended visit. MP
acknowledges the support from FONDECYT under the grant 1095027.

\end{document}